\documentclass[twocolumn,superscriptaddress]{revtex4}
\usepackage{empheq}
\usepackage{amsmath}
\usepackage{amssymb}
\usepackage{dsfont}
\usepackage{graphicx}
\usepackage{hyperref}
\usepackage{stackrel}
\usepackage{color}
\usepackage{comment}
\usepackage{slashed}

\def \ee{\end{equation}}
\def \be{\begin{equation}}
\def \eea{\end{eqnarray}}
\def \bea{\begin{eqnarray}}

\begin{document}

\title{Cosmological constant problem: deflation during inflation}

\author{Felipe Canales}
\affiliation{Pontificia Universidad Cat\'olica de Chile Instituto de F\'isica, Pontificia Universidad Cat\'olica de Chile, Casilla 306, Santiago, Chile}

\author{Benjamin Koch}
\affiliation{Pontificia Universidad Cat\'olica de Chile Instituto de F\'isica, Pontificia Universidad Cat\'olica de Chile, Casilla 306, Santiago, Chile}
\affiliation{Institut f\"ur Theoretische Physik,  Technische Universit\"at Wien,
 Wiedner Hauptstrasse 8-10, A-1040 Vienna, Austria}

\author{Cristobal Laporte}
\affiliation{Pontificia Universidad Cat\'olica de Chile Instituto de F\'isica, Pontificia Universidad Cat\'olica de Chile, Casilla 306, Santiago, Chile}

\author{ \'Angel Rinc{\'o}n }
\affiliation{Instituto de F\'isica, Pontificia Universidad Cat\'olica de Valpara\'iso, Avenida Brasil 2950, Casilla 4059, Valparaso, Chile.}
\email{angel.rincon@pucv.cl}

\begin{abstract}
We argue that the discrepancy between the Planck mass scale and the observed value of the cosmological constant can be largely attenuated if those quantities are understood as a result of effective, and thus scale-dependent, couplings.
We exemplify this mechanism for the early inflationary epoch of the universe by solving the corresponding effective gap equations, subject to an energy condition. 
Several non-trivial checks and extensions are discussed.
A comparison of our results to the renormalization group flow, obtained within the asymptotic safety program reveals a stunning agreement.
\end{abstract}

\maketitle

%\begin{keyword}
%\texttt{elsarticle.cls}\sep \LaTeX\sep Elsevier \sep template
%\MSC[2010] 00-01\sep  99-00
%\end{keyword}

%\end{frontmatter}

%\linenumbers
\tableofcontents
\section{Introduction}
The cosmological constant problem (CCP) seems to point towards deep
misconceptions in our current understanding of the interplay between quantum field theory and general relativity \cite{Adler:1995vd,Martel:1997vi,Weinberg:1988cp}.
In this article, we want to point out that one possible misconception is that,  one makes reference to gravitational couplings at different epochs of the universe,
in particular for the most dramatic version of the CCP.
This reference to different epochs of the Universe is however inconsistent if there is an epoch where the gravitational couplings are not constant, 
as it is expected for example from quantum
effects which are can induce an
effective running of the fundamental couplings  (e.g. Newtons coupling $G_0 \rightarrow G_k$).
It is clear that for quantum effects to be notable one needs a violent epoch of the universe. Due to our current understanding
the early inflationary period~\cite{Guth:1980zm,Linde:1981mu} is the best candidate for such non-classical corrections to be relevant.
In particular, during inflation the quantum effects and thus the scale-dependence of effective couplings
can be expected to be much stronger than during other later evolution periods.
There are two arguments that support this idea.
%,which is depicted in the schematic figure~\eqref{figscheme}.
First, since the very early universe is typically related to Planck-time and Planck-distances it is natural to expect important quantum gravity 
corrections~\cite{Mukhanov:1990me,Gasperini:1992em,Novello:2002nz,Ashtekar:2004eh,Thiemann:2002nj,Rovelli:2011eq,Wetterich:2017ixo,Rubio:2017gty},
such as scale-dependence of the effective couplings.
Second, 
in these very early times, other scales such as particle masses, which might moderate the scale-dependence are still irrelevant.
During later times, to the contrary, there are no Planck distances involved
and there are also logarithmically running particle masses contributing which could
 diminish the scale-dependence of the gravitational couplings further.
%
%\begin{figure}[ht!]
%\centering
%\includegraphics[width=\linewidth]{atcomic2.pdf}
%\caption{
%\label{figscheme} Schematic evolution of the scale factor $a(t)$ in the early universe.
%}
%\end{figure}
Thus, it makes sense to focus this type of study on the first inflationary period of the Universe.
Under such extreme conditions the gravitational couplings including the cosmological
``constant'' can not be expected to be constant \cite{Afshordi:2015iza} and thus one has to take
care when realizing a comparison of those couplings at different times.

In the following sections we will elaborate a scale-dependent vacuum dominated model which incorporates 
the following features, which are in agreement with the ideas mentioned above:
i) scale-dependence within effective gap equations
ii) inflationary behavior of the cosmological scale factor $a(t)$
iii) non--trivial evolution of the vacuum energy density and the gravitational coupling
and
iv) exponential suppression of the CCP during the inflationary period.
%
%{{ \bf
%\begin{itemize}
%\item scale-dependence within effective gap equations.
%\item Inflationary behavior of the cosmological scale factor $a(t)$.
%\item Non--trivial evolution of the vacuum energy density and the gravitational coupling.
%\item Exponential suppression of the CCP during the inflationary period.
%\end{itemize}
%}} \textcolor{cyan}{se piede resumir, lo reemplazo yo mismo}
%
This is achieved by
applying methods that have  been successfully used in order to include quantum correction into gravitational theories,
black hole backgrounds, wormholes, and particular cosmological problems \cite{Bonanno:2001xi,Reuter:2003ca,Reuter:2004nx,Weinberg:2009wa,Tye:2010an,Bonanno:2010bt,Eichhorn:2010tb,Litim:2011cp,Grande:2011xf,Benedetti:2012dx,Hindmarsh:2012rc,Eichhorn:2012va,Dona:2013qba,Copeland:2013vva,Bonanno:2015fga,Rodrigues:2015hba,Modesto:2015ozb,Rodrigues:2015rya,Rodrigues:2016tfm,Koch:2016uso,Platania:2017djo,Bonanno:2017gji,Rincon:2017ypd,Rincon:2017goj,Modesto:2017hzl,
Contreras:2017eza,Calmet:2017rxl,Toniato:2017wmk,Merzlikin:2017zan,Rincon:2018sgd,Hernandez-Arboleda:2018qdo,Contreras:2018dhs,Rincon:2018lyd,Rincon:2018dsq,Calmet:2018elv,Contreras:2018gct,Bonanno:2018gck,Contreras:2018gpl,Contreras:2018swc}.

%%%%%%%%%%%%%%%%%%%%%%%%%%%%%%%%
\subsection{Classical $\Lambda$ dominated universe}
%\label{subsecClU}
%
%\noindent{\bf{\em IA. Classical $\Lambda$ dominated universe.}}
%
Based on the cosmological principle and assuming a flat Universe, one uses the line element
\begin{equation} \label{metric}
\mathrm{d}s^2 = - \mathrm{d}t^2 + {a(t)}^2 
\Bigl[
%\frac{1}{1-\kappa r^2} 
\mathrm{d}r^2 
+ 
r^2 \mathrm{d}\Omega^2_2
\Bigl],
\end{equation}
as ansatz for the metric field, where $\mathrm{d}\Omega^2_2 = \mathrm{d}\theta^2 + \sin^2(\theta) \mathrm{d}\phi^2$.
The scale factor gives
\begin{equation}\label{classic_a_solution1}
    a(t)=a_i \text{e}^{t/\tau},
\end{equation}
where 
\be\label{deftau}
\tau = \pm \sqrt{\frac{3}{\Lambda_0 }}.
\ee
We will only use the positive sign in (\ref{deftau}),
which corresponds to an expanding universe.

%%%%%%%%%%%%%%%%%%%%%%%%%%%%%%%%%%%%%%%%%%%%
\subsection{The cosmological constant problem in the very early universe}
%%%%%%%%%%%%%%%%%%%%%%%%%%%%%%%%%%%%%%%%%%%%

\begin{figure}[ht!]
\centering
\includegraphics[width=\linewidth]{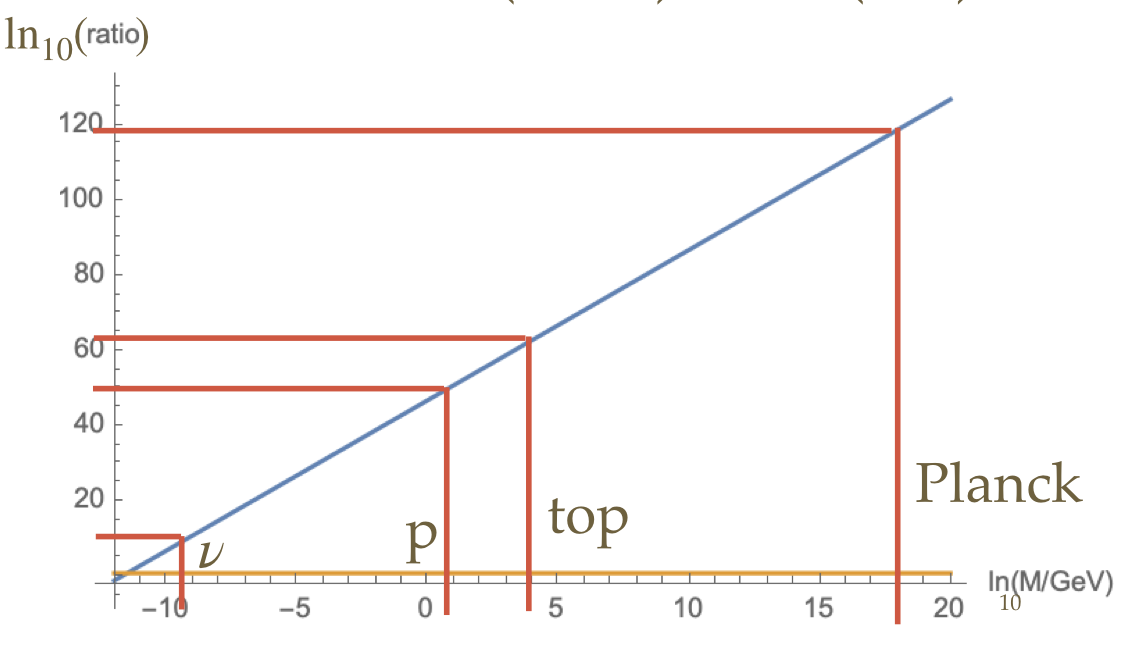}
\caption{
\label{figscales} The blue line indicates the logarithm of the dimensionless ratio (\ref{QFTratio}) as a function
of a cut-off mass scale $M_{UV}$.
A good agreement between the theoretical estimate and the observational value
is indicated by the orange line. 
}
\end{figure}

There are different versions of the CCP to be found in the literature.
Most of them can be cast in the dimensionless ratio between the highest UV cut-off $M_{UV}$ of
the quantum field theory one is using to calculate the vacuum energy density and
the energy density $\rho=\Lambda_0/G$ observed due to the accelerated expansion of the universe~\cite{Weinberg:1988cp}
\be\label{QFTratio}
r\approx \frac{M_{UV}^4}{\rho} \approx \frac{M_{UV}^4 G}{\Lambda_0}.
\ee
Figure \ref{figscales}  shows the exceedingly large values of (\ref{QFTratio})
for various emblematic energy scales $M_{UV}$ from neutrino physics, proton mass, top mass, up to 
the Planck scale.
For a good theoretical prediction, the value of (\ref{QFTratio}) is supposed to be
of order one, as indicated by the orange line.
Clearly, there is a dramatic discrepancy. 

As explained in the introduction, in this paper we will focus on the early expansion of the universe. During this epoch, 
 governed by pure gravity, the only dimensionful
quantity is actually the Planck mass given by $m_P^{-1}=\sqrt{8 \pi G}$.
If one uses this scale as UV scale $M_{UV}^2\approx 1/G$ to estimate the vacuum energy density,
the ratio (\ref{QFTratio}) simplifies to
\be\label{ratio}
r_{infl} \approx \frac{1}{G \cdot \Lambda_0}\approx 10^{122},
\ee
which is the line labeled with ``Planck'' in figure \ref{figscales}.
The fact that (\ref{ratio}) differs from one by more
than one hundred orders of magnitude is known as the cosmological constant problem in its worst form.
There have been numerous attempts to either solve, alleviate,
or to deny this problem \cite{Sahni:2002kh,Prokopec:2006yh,Brodsky:2009zd,Kaloper:2013zca,
Stojkovic:2014lha,Padilla:2015aaa,Bass:2015yaa,Novikov:2016hrc,Nojiri:2016mlb,Biemans:2016rvp,Wetterich:2017ixo,Hossenfelder:2018ikr}, but up to now the quarrel continues and no agreement has been achieved.

Since our purely gravitational approach ceases to be valid as soon as further mass scales such as the top mass
appear, we only can attempt to explain the discrepancy of 66 orders of magnitude spanned by the value of the ratio (\ref{ratio}) 
between the top mass and the Planck mass.

%%%%%%%%%%%%%%%%%%%%%%%%%%%%%%%%%%%%%%
\section{Scale--Dependence in Gravity}
%%%%%%%%%%%%%%%%%%%%%%%%%%%%%%%%%%%%%%
%
A very powerful tool in quantum field theory is the use of effective actions, which incorporate quantum effects in a scale--dependent modification of the classical action.
This type of scenario has been proposed and studied in the context of gravity in 
different ways and with different methods 
\cite{Weinberg:1976xy,Hawking:1979ig,Wetterich:1992yh,Morris:1993qb,Bonanno:2000ep,Litim:2002xm,Reuter:2004nv,Bonanno:2006eu,Niedermaier:2006ns,Percacci:2007sz}.
In the so-called Einstein-Hilbert truncation the effective action reads
%
%Promoting to scale-dependent couplings $G_{k}$ and $\Lambda_{k}$, the action is showed as
%
\begin{equation} \label{SD_EH_action}
\Gamma[g_{\mu\nu},k] = \int 
\mathrm{d}^4x \sqrt{-g} 
\Bigg[
\frac{1}{16 \pi G_{k} }\Bigl(R-2\Lambda_{k}\Bigl) \ + \ \mathcal{L}_M
\Bigg],
\end{equation}
where $\Lambda_k$ and $G_k$ are the cosmological and Newton scale--dependent couplings, respectively.
A great advantage of working with effective quantum actions like (\ref{SD_EH_action}) is
that they already incorporate the effects of quantum fluctuations like those 
being responsible for the CCP.
Thus, if the problem is properly addressed in terms of a background solution of this action,
 no additional quantum
corrections need to added.
In order to obtain background solutions for this effective action
one has to derive the corresponding gap equations.
Thus, varying the effective action respect to the inverse metric field, 
one obtains the corresponding Einstein field equations~\cite{Reuter:2004nv}
\begin{equation} \label{SD_EFE_vacuum}
G_{\mu\nu} =  T_{\mu \nu}^{\text{effec}}=- \Lambda(t) g_{\mu \nu} - {\Delta t}_{\mu\nu},
\end{equation}
where
\begin{equation} \label{delta_t_tensor}
{\Delta t}_{\mu\nu} = G(t)
\Bigl[
g_{\mu\nu}\nabla^{\alpha}\nabla_{\alpha}-\nabla_{\mu}\nabla_{\nu}
\Bigl]
G(t)^{-1}.
\end{equation}
In order to get physical information out of those equations
one has to set the renormalization scale in terms of the
physical variables of the system under consideration $k\rightarrow k(x, \dots)$. 

%%%%%%%%%%%%%%%%%%%%%%%%%%%%%%%
\subsection{Scale setting}
%\label{subsecSS}
%%%%%%%%%%%%%%%%%%%%%%%%%%%%%%%
%
For most types of scale settings \cite{Bonanno:1998ye,Bonanno:2000ep,Emoto:2005te,Bonanno:2006eu,Reuter:2006rg,Ward:2006vw,Koch:2007yt,Hewett:2007st,Litim:2007iu,Burschil:2009va,Falls:2010he,Casadio:2010fw,Basu:2010nf,Reuter:2010xb,Cai:2010zh,Falls:2012nd,Becker:2012js,Becker:2012jx,Koch:2013owa,Gonzalez:2015upa,Torres:2017ygl,Knorr:2017fus,Eichhorn:2017egq,Pawlowski:2018swz,Alkofer:2018zze,Eichhorn:2018yfc}
the identification $k\rightarrow k(x, \dots)$ breaks the reparametrization symmetry (\ref{SD_EFE_vacuum}).
There are, however, approaches which allow to maintain the symmetries
and thus the consistency of the system.
We will invoke a procedure which invokes the variational principle for~$k$
\be\label{VarSS}
\frac{\delta \Gamma_k}{\delta k}=0.
\ee
This approach
has been proposed in \cite{Koch:2010nn,Domazet:2012tw,Koch:2014joa} and
it is a straightforward implementation of background scale 
independence~\cite{Manrique:2009uh,Manrique:2010am,Manrique:2010mq,Becker:2014qya,Dietz:2015owa,Labus:2016lkh,Morris:2016spn,Ohta:2017dsq}.
Relation (\ref{VarSS}) is not a fixed point condition but rather a scale setting condition,
it even can be used when the renormalization group flow has already 
approached its fixed point \cite{Koch:2014cqa}.

Further, for highly symmetric systems with
a small number of couplings one can attempt to solve (\ref{SD_EFE_vacuum})
for $g_{\mu \nu}(x), \;G(x)$, and $\Lambda(x)$ directly by the use of 
a special ansatz \cite{Rincon:2017ypd,Rincon:2017goj,Rincon:2017ayr,
Contreras:2017eza,Rincon:2018sgd,Contreras:2018dhs,Rincon:2018lyd,Rincon:2018dsq,Contreras:2018gct} motivated
by the structure of the classical solution \cite{Jacobson:2007tj},
or by imposing an energy condition~\cite{Rincon:2017ayr}.

In this paper we will apply both strategies, the variational scale setting (\ref{VarSS}) and the energy 
condition mentioned above.
%
%
%%%%%%%%%%%%%%%%%%%%%%%%%%%%%%%%%%%%%%%%%%%%%
\subsection{Modified Friedmann equations}
%%%%%%%%%%%%%%%%%%%%%%%%%%%%%%%%%%%%%%%%%%%%%
%
Within the ansatz (\ref{metric}) the dynamical variable $a=a(t)$ does only depend on time and not on spatial directions,
the same will be assumed for the two additional variables $G=G(t)$ and $\Lambda=\Lambda(t)$.
With this, the gap equations (\ref{SD_EFE_vacuum})
take the form of generalized Friedmann equations
\begin{align} \label{SD_EFE_vacuum_00}
\left(\frac{\dot{a}}{a}\right)^2
- 
\frac{\Lambda }{3} 
&=
\frac{8 \pi G}{3} \rho(t),
\\
\label{SD_EFE_vacuum_11}
2 \frac{\ddot{a}}{a} 
+ 
\left(\frac{\dot{a}}{a}\right)^2 
-
\Lambda  
&=
-
 (8 \pi G)p(t),
%-
%2 \frac{a}{a^2} \left(\frac{{\dot{G}}}{G}\right)^2 
%+ 
%\frac{a}{a^2} \left(\frac{\ddot{G}}{G} \right)
%+ 
%2 \frac{\dot{a}}{a^2} \left(\frac{\dot{G}}{G}\right)
%
%
%\begin{equation} \label{SD_EFE_vacuum_11}
%\frac{2a(t)\ddot{a}(t)+\dot{a}(t)^2+\kappa }{{a(t)}^2}=\Lambda(t)+\frac{-2a(t){\dot{G}(t)}^2+a(t)G(t)\ddot{G}(t)+2G(t)\dot{a}(t)\dot{G}(t)}{{a(t)}^2{G(t)}^2}
\end{align}
where an effective density ($\rho$) and pressure ($p$), given by the scale--dependent framework have been identified
with the density and pressure parameters of a perfect fluid
\begin{align} 
\label{rhoeffec}
\frac{8 \pi G}{3} \rho(t) &\equiv \left(\frac{\dot{a}}{a}\right) 
\left(\frac{\dot{G}}{G}\right),
\\
\label{peffec}
-  (8 \pi G) p(t) &\equiv -
2 \left(\frac{{\dot{G}}}{G}\right)^2 
+ 
\left(\frac{\ddot{G}}{G} \right)
+ 
2 \left(\frac{\dot{a}}{a}\right) \left(\frac{\dot{G}}{G}\right).
\end{align}
Further, one has the scale setting relation (\ref{VarSS})
\be\label{VarSS2}
\frac{d}{dt}\Bigg[
\frac{1}{16 \pi G(t) }\Bigl(R-2\Lambda(t)\Bigl) 
\Bigg]=0.
\ee
There are three equations, namely the
two cosmological gap equations (\ref{SD_EFE_vacuum_00}, \ref{SD_EFE_vacuum_11}) and
the scale setting condition (\ref{VarSS2}). There are also  three unknown functions ($a(t),\, G(t),\, \Lambda(t)$).
However, the system respects general covariance, which implies that $\nabla_\mu G^{\mu \nu}=0$ and thus
one of the three equations (\ref{SD_EFE_vacuum_00}, \ref{SD_EFE_vacuum_11}, \ref{VarSS2}) 
is just a consequence of the other two.
This means that
one needs one additional relation or condition in order to determine a solution.
In the next subsection, we will invoke an energy condition in
order to provide the missing equation.

%{\bf{As anticipated in subsection \ref{subsecSS},
%this missing condition will be the explained in detail in the next subsection.}}\textcolor{cyan}{ pienso que se puede quitar.}

%%%%%%%%%%%%%%%%%%%%%%%%%%%%%%%%%%%%%%%%%%%%%%
%
\subsection{Null energy condition (NEC)}
%%%%%%%%%%%%%%%%%%%%%%%%%%%%%%%%%%%%%%%%%%%%%%
%
%\noindent{\bf{\em IIC. Null energy condition (NEC).}}
%
%%%%%%%%%%%%%%%%%%%%%%%%%%%%%%%%%%%%%%%%%%%%%55
In an empty universe, a light-like signal traveling on a null geodesic with velocity vector $\ell^\mu =d x^\mu/dt$ should not
perceive any energy density of the background.
This should be true for other vacuum contributions to the effective stress energy tensor, such as $\Delta t_{\mu \nu}$,
introduced by the scale-dependence of the gravitational coupling
\begin{equation}\label{NEC0}
T_{\mu \nu}^{\text{effec}} \ell^\mu \ell^\nu  = -\Delta t_{\mu\nu}\ell^\mu \ell^\nu  =0.
\end{equation}
This is reminiscent of the null energy condition \cite{Visser:1999de}.
The above relation is the scale setting condition which we will impose in addition to
the cosmological gap equations (\ref{SD_EFE_vacuum_00}, \ref{SD_EFE_vacuum_11}).
The vector field $\ell^{\mu}$ satisfying the geodesic equation can be written as 
$\ell^{\mu} = C_0 \ a^{-1} (1,(1-\kappa r^2)^{1/2} \ a^{-1},0,0).$
%
%\begin{equation}
%\ell^{\mu} = C_0 \ a^{-1} \Bigg(1, \Bigl(1-\kappa r^2 \Bigl)^{1/2} \ a^{-1},0,0 \Bigg).
%\end{equation}
%
Replacing this in  (\ref{NEC0}) gives the NEC condition for the cosmological model
\begin{equation} \label{NEC}
%\frac{{C_0}^2}{a^2 } 
%\Bigg( 
- 
2 \left(\frac{\dot{G}}{G}\right)^2
+
\left(\frac{\ddot{G}}{G} \right)
- 
\left(\frac{\dot{a}}{a}\right) \left(\frac{\dot{G}}{G} \right) 
%\Bigg) 
= 
0.
\end{equation}
%{\bf{This condition will be used in the following sections.}}\textcolor{cyan}{pienso que se puede quitar.}

%%%%%%%%%%%%%%%%%%%%%%%%%%%%%%%%
\section{Results and discussion}
%
%\noindent{\bf{\em III. Results and discussion.}}
%
%%%%%%%%%%%%%%%%%%%%%%%%%%%%%%%%%%%5
%
%
%{{\bf{In this section we will present the
%solution of the equations (\ref{SD_EFE_vacuum_00} , \ref{SD_EFE_vacuum_11}, and \ref{NEC}). 
%We will further analyze its properties and discuss its consequences, in particular on
%the CCP.}} \textcolor{cyan}{pienso que se puede quitar.}
%
%%%%%%%%%%%%%%
%\subsection{Solution}
%
The solution of  (\ref{SD_EFE_vacuum_00}), (\ref{SD_EFE_vacuum_11}), and (\ref{NEC}) 
that describes an expanding universe is
\begin{align}
\label{a_plano}
a(t) &=  a_i \text{e}^{\frac{t}{\tau}},
\\
\label{G_plano}
G(t) &=G_0 \Bigg[1+ \tilde \xi \displaystyle\int_{t_0}^{t} a(t')dt' \Bigg]^{-1}= \frac{G_0}{1 + \xi a(t)},
\\
\label{L_plano}
\Lambda(t) &= \Lambda_0 
\Bigg[
\frac{ 1 + 2\xi a(t)}{1 + \xi a(t)}
\Bigg].
\end{align}
It is direct to see that  (\ref{a_plano}) takes the same form as the classical counterpart
(\ref{classic_a_solution1}).
A generalization to non zero values of $\kappa$ is straight 
forward (see version one of \cite{Canales:2018tbn}).
The solution involves four integration constants which we have labeled 
$(a_i, G_0, \Lambda_0, \xi)$.
They are set by the conditions
\bea
\lim_{t\to 0}a(t) = a_i,
\hspace{0.6cm}
\lim_{\xi\to 0}G(t) = G_0,
\hspace{0.6cm}
\lim_{\xi\to 0}\Lambda(t) = \Lambda_0.
\eea
One notes that the parameter $\xi$ plays the role of a control parameter,
When setting $\xi$ to zero, the time-dependent couplings $G(t)$ and $\Lambda(t)$ are reduced to their
constant and thus ``classical'' counterparts ($G_0$ and $\Lambda_0$).
Please note that the expansion rate is given by $\tau = \pm \sqrt{3/\Lambda_0 }$,
which looks exactly like the classical rate (\ref{deftau}). There is however an important 
conceptual difference
since now, $\Lambda_0$ is one of the integration constants, while in (\ref{deftau}) it
was given as input parameter.

%
%
%
%%%%%%%%%%%%%%%%%%%%%%%%%%%%%%%%%%%%%%%%%%%%%%%%%%%%%%%%%%%%%%%%%%%%%
%\begin{figure}[ht!]
%\centering
%\includegraphics[width=\linewidth]{GLt_Angel.pdf}
%%\includegraphics[width=0.48\linewidth]{GLt_Angel.pdf}
%%\
%%\includegraphics[width=0.48\linewidth]{GLt_Angel.pdf}
%\caption{
%		\label{fig:2}
%		Evolution of $G(t)\cdot \Lambda(t)$
% for $\xi=0$ 
%		(solid 
%		black 
%		line), $\xi=0.001$ 
%		(dotted--dashed 
%		red 
%		line), $\xi = 0.01$  
%		(dotted 
%		green 
%		line) and $\xi=0.1$ 
%		(dashed 
%		blue 
%		line).
%The rest of parameters have been taken to be unity.
%}
%\end{figure}
%%
%The behavior of the scale-dependent couplings, in particular the product $G\cdot \Lambda$
%as a function of $t$ is exemplified in the figure \ref{fig:2}.
%One notes that $G(t)\cdot \Lambda(t)$ experiences a strong suppression. One further notes that
%for smaller values of $\xi$ the evolution of the couplings
%starts at later times. Apart from this change, the functions are form-invariant under $t\rightarrow t-t_i$ transformations. 
%This means that the value of $\xi$ can be reabsorbed in a shift in the initial or
%final conditions for $a(t)$, $G(t)$, and $\Lambda(t)$.
%%%%%%%%%%%%%%%%%%%%%%%%%%%%%%%%%%%%%%%%%%%%%%%%%%%%%%%%%%%%%%%%%%%%%
%
%%%%%%%%%%%%%%%%%%%%%%%%%%%%%%%%%%%%%%%%%%%%%%%%%%%%%%%%
\subsection{Equation of state of scale-dependent $G(t)$}
%
%\noindent{\bf{\em IIIA. Equation of state of scale-dependent $G(t)$.}}
%
%%%%%%%%%%%%%%%%%%%%%%%%%%%%%%%%%%%%%%%%%%%%%%%%%%%%%%%%%%%%5
%
%{{\bf It is interesting to revisit a vacuum solution of the Einstein field equations ($T^{M}_{\mu \nu}=0$) where one observes that the inclusion of a 
%running of the gravitational coupling can be labeled as effective density (\ref{rhoeffec}) and pressure (\ref{peffec}) both arising from $\Delta t_{\mu \nu}$. }} \textcolor{cyan}{pienso que se puede quitar.}
%
Is there a meaningful equation of state $p=\omega \rho$
%\be
%p=\omega \rho
%\ee
for the effective type of pressure and energy density?
Subtracting (\ref{SD_EFE_vacuum_11}) from $3\times$(\ref{SD_EFE_vacuum_00}),
one finds for the solution (\ref{a_plano}, \ref{G_plano})
\be
p(t)+\rho(t)=0.
\ee
Thus, the effective parameters take the form of a ``classical" Dark Energy equation of state with $\omega = -1$.
Thus, the effective pressure and the effective energy density cancel each other, which explains why the scale--dependence parameter $\xi$ does not appear in the solution for $a(t)$.

%%%%%%%%%%%%%%%%%%%%%%%%%%%%%%%%%%%%%%%%%%%%
\subsection{Deflation of the cosmological constant problem}
%%%%%%%%%%%%%%%%%%%%%%%%%%%%%%%%%%%%%%%%%%%%%%%%%%%%%%%%%%%%5
%
%\noindent{\bf{\em IIIB. Deflation of the CCP.}}
%
%%%%%%%%%%%%%%%%%%%%%%%%%%%%%%%%%%%%%%%%%%%%%%%%%%%%%%%%%%%%5
As explained in the introduction, the CCP 
in a Planck era universe can be parametrized by
the value of $\Lambda \cdot G$ (\ref{ratio}).
However, the quantum gravity calculation makes reference
to the very early Planck era $t_i$, while the measurement makes reference
to the more recent history of the universe, long after the end of inflation $t_f$.
scale-dependence can offer alleviation of this problem since it
allows for a continuous evolution of $\Lambda(t) \cdot G(t)$
from the Planck era $t_i$ to the end of inflation $t_f$ as shown in the left panel of figure \ref{fig:2}.
As argued in the introduction, we assume that the most dominant scaling effects
occur in this period and that in later times, after $t_f$, the 
scaling effects will be dominated by coupling to 
external particle scales $m_i$, which only run logarithmically. 

In order to show that the CCP can be addressed effectively within our framework,
we impose six conditions
on the scale-dependent solution (\ref{a_plano}-\ref{L_plano}) during inflation.
One has to check whether those conditions
can be met by a choice of the constants $t_i, t_f, G_0,\, \Lambda_0,\, a_i,\, \xi$.

\begin{figure}[ht!]
\centering
\includegraphics[width=0.8\linewidth]{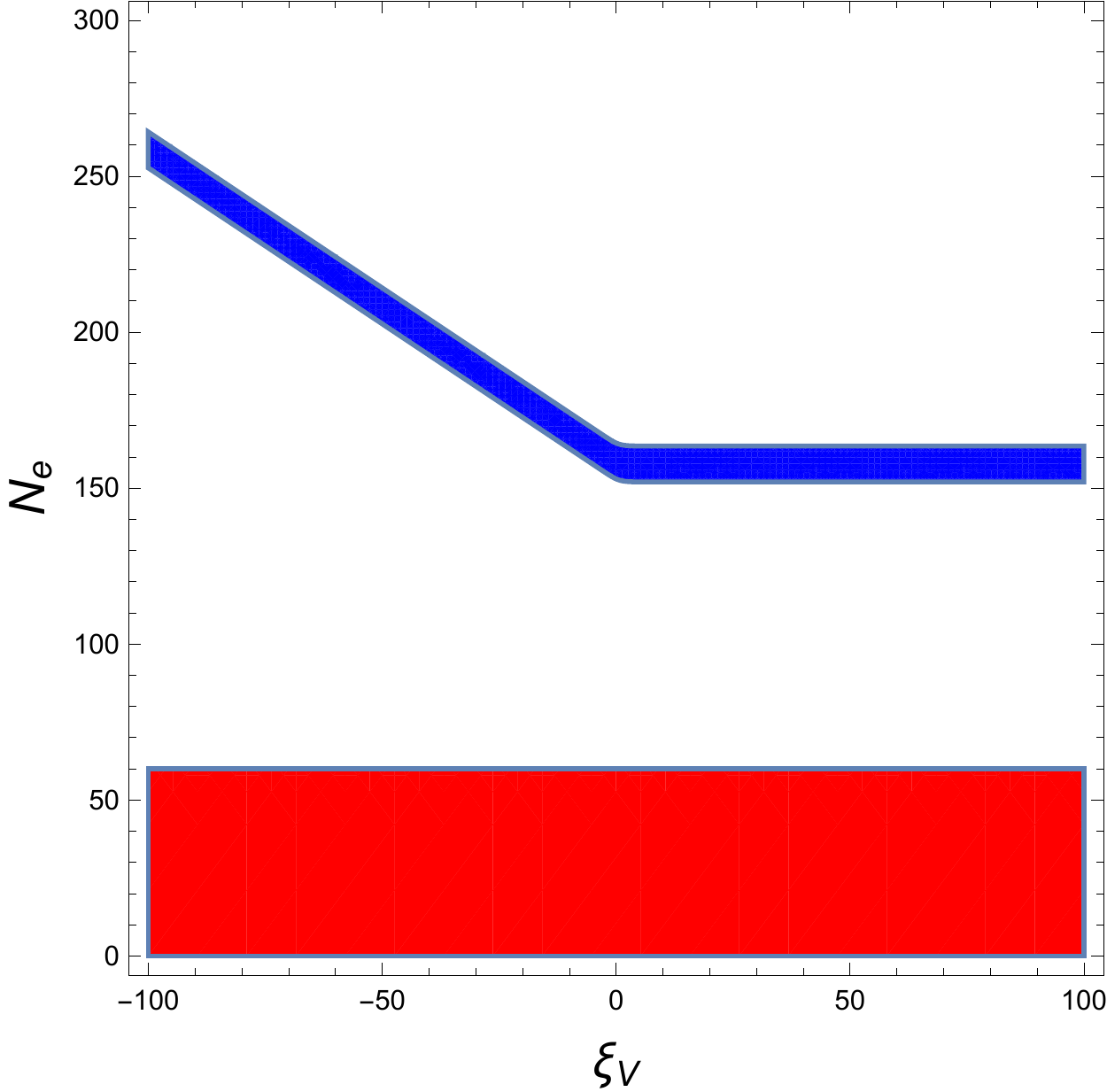}
\caption{
		\label{figNee}$N_e$, $\xi_V$ parameter space, where
        the blue contour is fulfilling both conditions of 					(\ref{inequ}) and the red colour is 				excluded from the observational condition  (\ref{Neobs}).
}
\end{figure}

\begin{itemize}
\item[a)] At the initial time $t_i\equiv 0$ one demands
\be\label{ai}
a(t_i)=1.
\ee
and
\be\label{LGt0}
\Lambda(t_i)\cdot G(t_i)\approx 1,
\ee
which meets the theoretical expectation.
\item[b)]
At the end of inflation $t_f$ one imposes that 
the gravitational coupling has evolved to a value 
close to the current measured value
\be\label{Gtf}
G(t_f)\approx G_N ,
\ee
and that the cosmological constant problem has gravitationally evolved to
\be\label{LGtf}
\Lambda(t_f)\cdot G(t_f)\approx 5 \cdot 10^{-66},
\ee
around the magnitude dictated by the heaviest standard
model particle.
\item[c)]
The observed flatness of the universe suggest further that 
during this epoch (from $t_i$ to $t_f$) of rapid growth, the universe expanded at least
sixty e-folds~\cite{Patrignani:2016xqp}
\be\label{Neobs}
N_e \ge 60.
\ee
\end{itemize}
By imposing the conditions (\ref{ai}, \ref{LGt0}, \ref{Gtf}) one finds
\bea
a_i &=&1,
\\ 
t_f&=& N_e \tau,
%t_f&=& N_e \sqrt{\frac{3}{\Lambda_0}},
\\
\Lambda_0&=&\frac{(1+\xi)^2}{G_N (1+2 \xi)\left(1+\xi e^{N_e} \right)}.
\eea
In order to analyze the condition (\ref{LGtf}), we put it into an order of magnitude inequality, accepting
five orders of magnitude deviation from the value given in (\ref{LGtf})
\be\label{inequ}
 10^{-71}\lessapprox \Lambda(t_f)\cdot G(t_f)\lessapprox 10^{-61}.
\ee
This double inequality puts conditions on the two remaining constants $N_e$ and $\xi$.
The region of allowed values for those two constants, according to (\ref{inequ}), is shown in the blue contour in figure \ref{figNee},
where we have defined $\xi= e^{\xi_V}$.
%
%\begin{figure}[h!]
%        \centering
%        \includegraphics[width=0.85\textwidth,height=0.35\textheight]{Ne.pdf}
%        \caption{\label{figNee}. $N_e$, $\xi_V$ parameter space, where
%        the blue contour is fulfilling both conditions of (\ref{inequ}) and the red condour
%        is excluded from the observational condition  (\ref{Neobs}).
%        }
%\end{figure}
From this figure one notes that the condition (\ref{inequ}) never gets in conflict with the observational bound (\ref{Neobs}).
%Arbitrarily small values of $\xi$ are allowed if one allows for very large value of $N_e$.
It is thus shown that substantial alleviation of the cosmological constant problem within the presented framework can be achieved. For this one
does need the inflationary period to last at least $\ln(10^{66})\approx 150$ e-folds.
This is significantly larger than the experimental lower bound, but it does not cause tension
with observational data since it is
very hard to establish a distinctive upper bound for this quantity \cite{Banks:2003pt}.

The scale-dependent evolution could be incorporated
into more realistic models of inflation~\cite{Kolb:1990vq}, but this goes beyond the scope of this article.
It is further tempting to speculate whether the suppression of $G(t)$ in the
first equality of (\ref{G_plano}) can be used in combination 
with the scale-dependence of other couplings in the post-inflationary universe~\cite{Capozziello:1996xq,Calmet:2001nu,Carneiro:2003as,Reisenegger:2009cq,Yunes:2009bv,Uzan:2010pm,Fritzsch:2010zzc,Anderson:2015bva,Kantha:2016ylw,Calmet:2017czo}
in order to contribute to the solution
of the hierarchy problem~\cite{Martin:1997ns,ArkaniHamed:1998rs,Meissner:2006zh}. This possibility will be explored in future studies.

%%%%%%%%%%%%%%%%%%%%%%%%%
\subsection{Comparison with the functional renormalization group}
%%%%%%%%%%%%%%%%%%%%%%%%%
%
%\noindent{\bf{\em IIIC. Comparison with the functional renormalization group.}}
%
Even though our calculation never makes use of specific quantum gravity
beta functions, it is tempting to make a comparison with the findings in such an approach.
The discussion below will be complementary to the findings in \cite{Wetterich:2017ixo,Rubio:2017gty},
where strong effects of the evolution of couplings on the cosmological constant problem have been pointed out in context with a possible infra-red instability of asymptotically save gravity~\cite{Houthoff:2017oam,Bosma:2019aiu}

Using functional renormalization group (RG) methods, in particular,
the Wetterich equation~\cite{Wetterich:1992yh,Morris:1993qb,Litim:2002xm}, it is possible to obtain
a prediction for the scale-dependent couplings in gravitation actions such as (\ref{SD_EH_action}).
The dimensionless couplings $\hat g(k)=\hat G(k)k^2$ and $\hat \lambda(k)=\hat \Lambda(k)/k^2$ are found for the dimensionless renormalization scale
$\hat t=\log(k/k_0)$
\bea\label{gAS}
\hat g(\hat t)&=& \frac{g_0 e^{2 \hat t}}{1+g_0\left(e^{2 \hat t}-1\right)/g^*},\\ \label{lAS}
\hat\lambda(\hat t)&=& \frac{g^* \lambda_0 + e^{-2 \hat t}\left(e^{4 \hat t} -1\right)g_0 \lambda^* }{1+g_0\left(e^{2 \hat t}-1\right)/g^*},
\eea
where $g^*$ and $\lambda^*$ are the values of the ultra violet fixed point
predicted by the Asymptotic Safety (AS) conjecture~\cite{Weinberg:1976xy,Percacci:2007sz} and found within the functional RG approach~\cite{Reuter:1996cp,Reuter:2001ag,Litim:2003vp}.
Further,
$\lambda_0$ and $g_0$ are initial values defined at an intermediate renormalization scale $\hat t=0$.
For the CCP in the very early universe the relevant quantity is the product of the dimensionful couplings $\hat G(\hat t)\cdot \hat \Lambda(\hat t)$.
In four dimensions the 
product of the dimensionful couplings is equal to the product of the dimensionless couplings (\ref{gAS}, \ref{lAS})
\be\label{GLAS}
\hat G(\hat t)\hat\Lambda(\hat t)=g(\hat t) \lambda(\hat t).
\ee
Therefore, a comparison of the naturally dimensionless product (\ref{GLAS}) allows for much more robust statements
than a comparison of the individual dimensionful or dimensionless couplings.

\begin{figure}[ht!]
\centering
\includegraphics[width=0.48\linewidth]{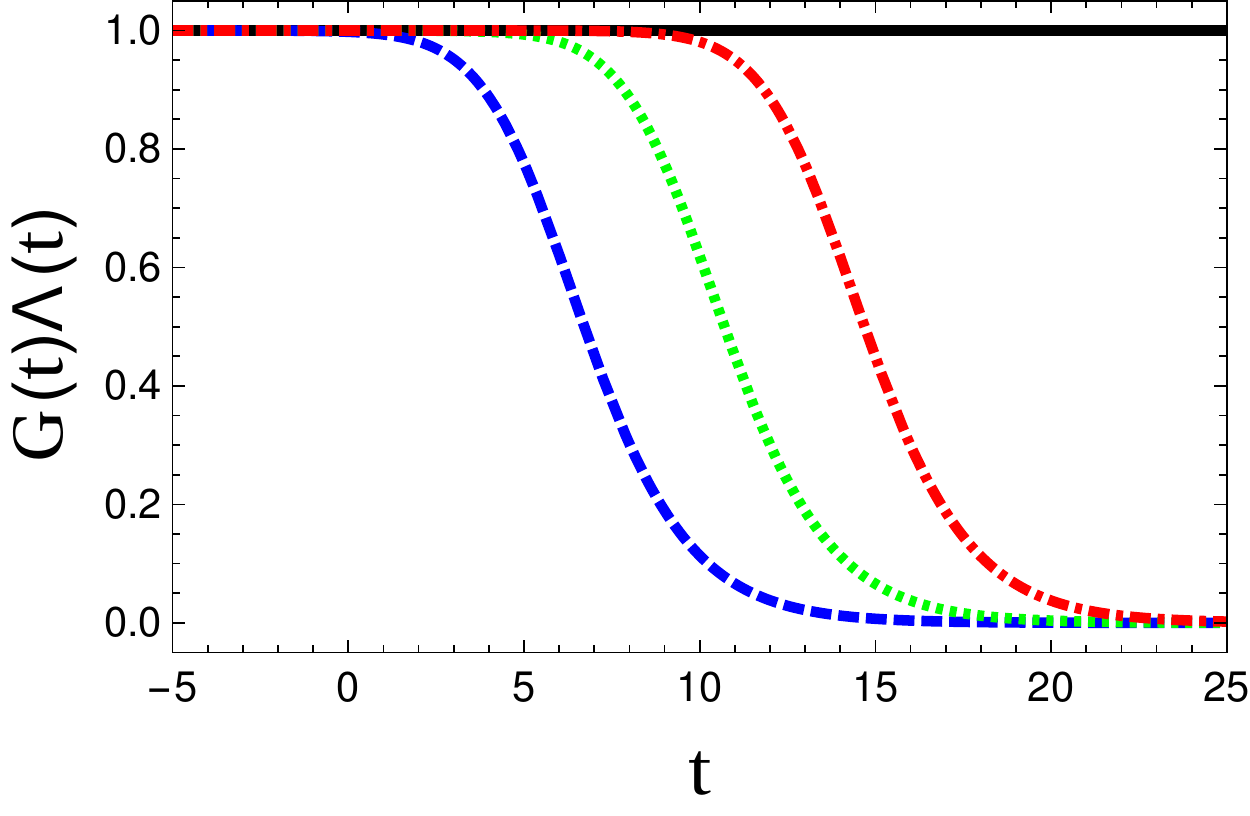}
\
\includegraphics[width=0.48\linewidth]{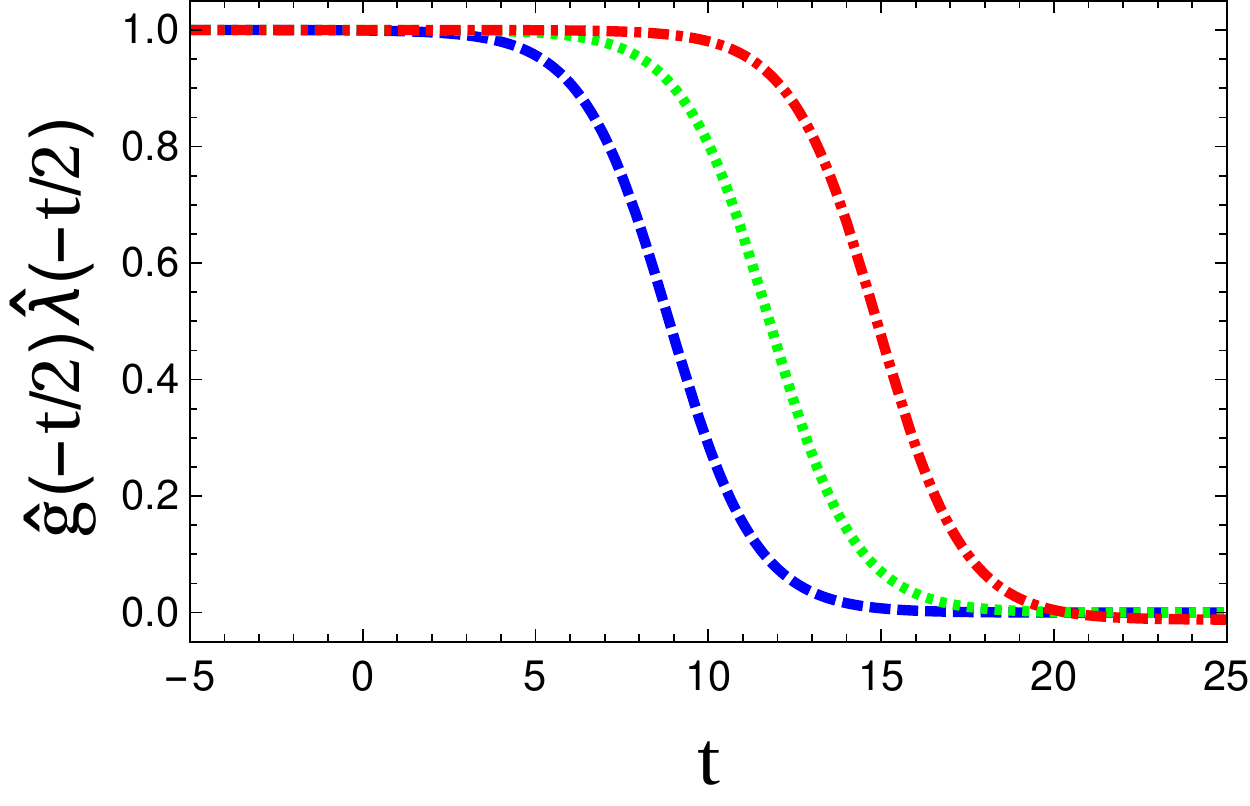}
\caption{
		\label{fig:2}
		LEFT: Evolution of $G(t)\cdot \Lambda(t)$
 	for $\xi=0$ 
		(solid 
		black 
		line), $\xi=0.001$ 
		(dotted--dashed 
		red 
		line), $\xi = 0.01$  
		(dotted 
		green 
		line) and $\xi=0.1$ 
		(dashed 
		blue 
		line).
	The rest of the parameters have been taken to be unity.
	RIGHT: Evolution of  $\hat{\lambda}(-t/2) \cdot \hat{g}(-t/2)$ as given by AS 	for $g_0 =0.001$ (dashed blue line), for $g_0 = 0.01$ (dotted green line), 	and for $g_0 = 0.1$ (dot-dashed red line). Other constants were
	set to unity.
	}
\end{figure}

The one million dollar question is whether there is a scale setting $\hat t=\hat t(t)$ for which the product (\ref{GLAS}) obtained from AS
agrees, at least qualitatively, with the product $G(t)\cdot \Lambda(t)$ derived in our scale-dependent cosmological solution (\ref{G_plano}, \ref{L_plano}).
In the ultraviolet (UV) and close to the separatrix~\footnote{With ``separatrix'' it is referred 
to the RG flow line
which connects the trivial fixed point $g=\lambda=0$ with the non trivial UV fixed point $(g^*,\,\lambda^*)$.} of the RG  flow one can approximate (\ref{gAS} and \ref{lAS}) for
$e^{2\hat t}, g^*/g_0\gg1$, which gives
%\be
%\hat g(\hat t) \hat \lambda(\hat t)= g^* \lambda^*
%\frac{\frac{g^* \lambda_0}{g_0 \lambda^*}+e^{2 \hat t}}{\left( e^{2 \hat t}+\frac{g^*}{g_0}\right)^2}.
%\ee
%
\be
\hat g(\hat t) \hat \lambda(\hat t) = g^* \lambda^*
\left( 
\frac{g^* \lambda_0}{g_0 \lambda^*}+e^{2 \hat t}
\right)
\left( e^{2 \hat t}+\frac{g^*}{g_0}\right)^{-2} .
\ee
By inspection one realizes that the simple replacements
%%%%\bea\label{repl1}
%%%% g^* \lambda^*&\rightarrow& G_0 \Lambda_0, \\ 
%%%% \label{repl2}
%%%%g_0 &\rightarrow& \frac{G_0}{a_i \xi},\\ 
%%%%\label{repl3}
%%%%\hat t &\rightarrow& -\frac{t}{2 \tau},
%%%%\eea
%
%
\bea \label{identifi}
 g^* \lambda^*\rightarrow G_0 \Lambda_0,
\hspace{0.4cm}
g_0 \rightarrow \frac{G_0}{a_i \xi},
\hspace{0.4cm}
\hat t \rightarrow -\frac{t}{2 \tau},
\eea
together with the choice $\lambda_0=2 \lambda^*$, leads to a stunning exact matching
\be\label{id}
\left.\hat g(\hat t) \hat \lambda(\hat t)\right |^{\mbox{}}_{\mbox{(3.19)}}=G(t) \Lambda(t).
\ee
This means that for a proper scale setting and  a proper choice
of the flow trajectory, both approaches
show the same type of deflation of the cosmological constant problem.
Graphically this behavior is shown in figure \ref{fig:2}.

\noindent Comparing our solutions to the AS predictions, on one hand, we observe that the behavior of the product $G\cdot \Lambda$ as a function of $t$ is exemplified in the figure \eqref{fig:2} (LEFT). One notes that $G(t)\cdot \Lambda(t)$ experiences a strong suppression. In addition, for smaller values of $\xi$ the evolution of the couplings starts at later times. Apart from this change, the functions are form-invariant under $t\rightarrow t-t_i$ transformations. This means that the value of $\xi$ can be reabsorbed in a shift in the initial or final conditions for $a(t)$, $G(t)$, and $\Lambda(t)$. 
On the other hand, the same features are shown in \eqref{fig:2} (RIGHT).

Clearly, this agreement can still be a coincidence but it appears sufficiently
interesting to us to mention it here in the discussion. 
There are several comments about the result (\ref{id}) and the identifications 
\eqref{identifi}
%
%(\ref{repl1}-\ref{repl3})
%
to be made.
The result (\ref{id}) is not the product of a 
very particular cut-off identification. It is straightforward to check that the same agreement can
be obtained from different beta functions arising from different cut-off prescriptions.
From  \eqref{identifi}
%(\ref{repl1}) 
one sees from the UV/AS perspective that the control parameter $\xi$
is not necessarily small, but a rather large number.
This means that scale-dependence in this regime of high energies is a dominant effect and not just a small correction to classical behavior. 
One further sees from \eqref{identifi}
that the product  $\Lambda\cdot  G$ in the UV is predicted to
be just the product of the two fixed points $g^* \cdot \lambda^*$ of the RG flow.
The values of those fixed points
are typically found to be of order$\sim 0.2$, the product is thus not dramatically far from
unity, just as we had supposed 
in our working assumptions.
Finally, one notes from \eqref{identifi}
that the scale identification 
is in agreement with 
our intuitive expectations, since it associates very early times $t\rightarrow -\infty$ to very 
large RG scales $k\rightarrow \infty$ and late times $t\rightarrow \infty$ 
to small RG scale $k\rightarrow 0$.

We can not resist to insist that the identity (\ref{id})
is absolutely not trivial. It 
links two completely different and independent
calculations.
One is implementing quantum effects into a gravitational theory by
means of effective quantum field theory and the other
solving the modified Friedmann  equations (\ref{SD_EFE_vacuum_00}, \ref{SD_EFE_vacuum_11})  subject to a simple energy condition (\ref{NEC}).
The support of both approaches is mutual.
 One can read the agreement of (\ref{G_plano}, \ref{L_plano})
with (\ref{gAS}, \ref{lAS}) as support of the approach and the solution presented in this paper.
One can also conclude that AS is in good
shape to contribute to the solution of the CCP, if one chooses
a flow line which is in sufficiently close vicinity of the separatrix line ($g_0\ll 1$).

%%%%%%%%%%%%%%%%%%%%%%%%%%
\section{Conclusions}
%
%%%%%%%%%%%%%%%%%%%%%%%%%%
%
%\noindent{\bf{\em IV. Conclusions.}}
%
In this article, we presented a study on the early inflationary universe in the context of scale-dependence.
We give reasons, why it can be expected that the leading gravitational couplings experience scale-, in particular time-dependence
in this early epoch.
It is then argued that, if one accepts possible scale-dependence of the couplings, then
the CCP is ill-defined since it makes an identification of the product
$\Lambda(t) \cdot G(t)$ at two different times of the cosmological evolution. We conjecture that
 this misconception and thus the CCP can be resolved if
 one takes into account a proper scale-dependence of those couplings.

As a concrete example, we solve the gap equations (\ref{SD_EFE_vacuum}) of the scale-dependent gravitational action in the Einstein-Hilbert truncation
for the homogenous isotropic line element (\ref{metric}) subject to a null energy condition (\ref{NEC0}).
It is found that within this solution (\ref{a_plano}-\ref{L_plano}) 
the scale factor $a(t)$ grows exponentially. 
Further, the couplings $\Lambda(t)$ and $G(t)$ have the expected 
strong time-dependence. 
It is shown that the time evolution of $\Lambda(t) \cdot G(t)$ indeed allows accommodating
a substantial alleviation of the CCP in agreement with the conditions (\ref{ai}-\ref{LGtf}) and the observational
lower bound (\ref{Neobs}).

We note two particular features of our results.
First, the contribution of the scale-dependent coupling $G(t)$,
can be written as an ideal fluid obeying an equation of state, which is usually associated with vacuum energy density.
This is a strong support for the chosen energy condition (\ref{NEC0}), because it shows
that the condition guaranties a vacuum type equation of state for the scale dependent contribution $\Delta t_{\mu \nu}$
of the stress energy tensor.
Second, a comparison between the product $\Lambda(t) \cdot G(t)$, deduced in this paper, with
the product $\hat \lambda(\hat t) \cdot \hat g(\hat t)$, obtained from functional RG methods
within the AS scenario, reveals highly non-trivial similarities.

%%%%%%%%%%%%%%%%%%%%%%%%%%%%%%%%%%%%%%%%%%%%%%%%%%%%%%%%%%%%%%%%%%%%%%%%%%%%%%%%%%%%%%%%%%%%%%%%%%%%%%%%
%
%\section*{Acknowlegements}
%
\noindent{\bf{\em Acknowlegements.}}
B.~K. was supported by the Fondecyt 1161150 and Fondecyt 1181694. 
\'A. ~R. acknowldeges DI-VRIEA for financial support through Proyecto Postdoctorado 2019 VRIEA-PUCV.

%%%%%%%%%%%%%%%%%%%%%%%%%%%%%%%%%%%%%%%%%%%%%%%%%%%%%%%%%%%%%%%%%%%%%%%%%%%%%%%%%%%%%%%%%%%%%%%%%%%%%%%%


\begin{thebibliography}{99}

%\cite{Adler:1995vd}
\bibitem{Adler:1995vd} 
  R.~J.~Adler, B.~Casey and O.~C.~Jacob,
  %``Vacuum catastrophe: An Elementary exposition of the cosmological constant problem,''
  Am.\ J.\ Phys.\  {\bf 63}, 620 (1995).
  doi:10.1119/1.17850
  %%CITATION = doi:10.1119/1.17850;%%
  %20 citations counted in INSPIRE as of 15 Apr 2019


%\cite{Martel:1997vi}
\bibitem{Martel:1997vi} 
  H.~Martel, P.~R.~Shapiro and S.~Weinberg,
  %``Likely values of the cosmological constant,''
  Astrophys.\ J.\  {\bf 492}, 29 (1998)
  doi:10.1086/305016
  [astro-ph/9701099].
  %%CITATION = doi:10.1086/305016;%%
  %234 citations counted in INSPIRE as of 15 Apr 2019


%\cite{Weinberg:1988cp}
\bibitem{Weinberg:1988cp} 
  S.~Weinberg,
  %``The Cosmological Constant Problem,''
  Rev.\ Mod.\ Phys.\  {\bf 61}, 1 (1989).
  doi:10.1103/RevModPhys.61.1
  %%CITATION = doi:10.1103/RevModPhys.61.1;%%
  %4155 citations counted in INSPIRE as of 15 Apr 2019


%\cite{Guth:1980zm}
\bibitem{Guth:1980zm} 
  A.~H.~Guth,
  %``The Inflationary Universe: A Possible Solution to the Horizon and Flatness Problems,''
  Phys.\ Rev.\ D {\bf 23}, 347 (1981)
  [Adv.\ Ser.\ Astrophys.\ Cosmol.\  {\bf 3}, 139 (1987)].
  doi:10.1103/PhysRevD.23.347
  %%CITATION = doi:10.1103/PhysRevD.23.347;%%
  %7233 citations counted in INSPIRE as of 15 Apr 2019


%\cite{Linde:1981mu}
\bibitem{Linde:1981mu} 
  A.~D.~Linde,
  %``A New Inflationary Universe Scenario: A Possible Solution of the Horizon, Flatness, Homogeneity, Isotropy and Primordial Monopole Problems,''
  Phys.\ Lett.\  {\bf 108B}, 389 (1982)
  [Adv.\ Ser.\ Astrophys.\ Cosmol.\  {\bf 3}, 149 (1987)].
  doi:10.1016/0370-2693(82)91219-9
  %%CITATION = doi:10.1016/0370-2693(82)91219-9;%%
  %4499 citations counted in INSPIRE as of 15 Apr 2019


%\cite{Mukhanov:1990me}
\bibitem{Mukhanov:1990me} 
  V.~F.~Mukhanov, H.~A.~Feldman and R.~H.~Brandenberger,
  %``Theory of cosmological perturbations. Part 1. Classical perturbations. Part 2. Quantum theory of perturbations. Part 3. Extensions,''
  Phys.\ Rept.\  {\bf 215}, 203 (1992).
  doi:10.1016/0370-1573(92)90044-Z
  %%CITATION = doi:10.1016/0370-1573(92)90044-Z;%%
  %2745 citations counted in INSPIRE as of 15 Apr 2019


%\cite{Gasperini:1992em}
\bibitem{Gasperini:1992em} 
  M.~Gasperini and G.~Veneziano,
  %``Pre - big bang in string cosmology,''
  Astropart.\ Phys.\  {\bf 1}, 317 (1993)
  doi:10.1016/0927-6505(93)90017-8
  [hep-th/9211021].
  %%CITATION = doi:10.1016/0927-6505(93)90017-8;%%
  %848 citations counted in INSPIRE as of 15 Apr 2019


%\cite{Novello:2002nz}
\bibitem{Novello:2002nz} 
  M.~Novello, J.~Barcelos-Neto and J.~M.~Salim,
  %``A Model for time-dependent cosmological constant and its consistency with the present Friedmann universe,''
  Class.\ Quant.\ Grav.\  {\bf 19}, 3107 (2002)
  doi:10.1088/0264-9381/19/11/323
  [hep-th/0202169].
  %%CITATION = doi:10.1088/0264-9381/19/11/323;%%
  %6 citations counted in INSPIRE as of 15 Apr 2019


%\cite{Ashtekar:2004eh}
\bibitem{Ashtekar:2004eh} 
  A.~Ashtekar and J.~Lewandowski,
  %``Background independent quantum gravity: A Status report,''
  Class.\ Quant.\ Grav.\  {\bf 21}, R53 (2004)
  doi:10.1088/0264-9381/21/15/R01
  [gr-qc/0404018].
  %%CITATION = doi:10.1088/0264-9381/21/15/R01;%%
  %1264 citations counted in INSPIRE as of 15 Apr 2019


%\cite{Thiemann:2002nj}
\bibitem{Thiemann:2002nj} 
  T.~Thiemann,
  %``Lectures on loop quantum gravity,''
  Lect.\ Notes Phys.\  {\bf 631}, 41 (2003)
  doi:10.1007/978-3-540-45230-0 3
  [gr-qc/0210094].
  %%CITATION = doi:10.1007/978-3-540-45230-0 3;%%
  %332 citations counted in INSPIRE as of 15 Apr 2019


%\cite{Rovelli:2011eq}
\bibitem{Rovelli:2011eq} 
  C.~Rovelli,
  %``Zakopane lectures on loop gravity,''
  PoS QGQGS {\bf 2011}, 003 (2011)
  doi:10.22323/1.140.0003
  [arXiv:1102.3660 [gr-qc]].
  %%CITATION = doi:10.22323/1.140.0003;%%
  %216 citations counted in INSPIRE as of 15 Apr 2019


%\cite{Wetterich:2017ixo}
\bibitem{Wetterich:2017ixo} 
  C.~Wetterich,
  %``Graviton fluctuations erase the cosmological constant,''
  Phys.\ Lett.\ B {\bf 773}, 6 (2017)
  doi:10.1016/j.physletb.2017.08.002
  [arXiv:1704.08040 [gr-qc]].
  %%CITATION = doi:10.1016/j.physletb.2017.08.002;%%
  %26 citations counted in INSPIRE as of 15 Apr 2019


%\cite{Rubio:2017gty}
\bibitem{Rubio:2017gty} 
  J.~Rubio and C.~Wetterich,
  %``Emergent scale symmetry: Connecting inflation and dark energy,''
  Phys.\ Rev.\ D {\bf 96}, no. 6, 063509 (2017)
  doi:10.1103/PhysRevD.96.063509
  [arXiv:1705.00552 [gr-qc]].
  %%CITATION = doi:10.1103/PhysRevD.96.063509;%%
  %28 citations counted in INSPIRE as of 15 Apr 2019


%\cite{Houthoff:2017oam}
\bibitem{Houthoff:2017oam} 
  W.~B.~Houthoff, A.~Kurov and F.~Saueressig,
  %``Impact of topology in foliated Quantum Einstein Gravity,''
  Eur.\ Phys.\ J.\ C {\bf 77}, 491 (2017)
  doi:10.1140/epjc/s10052-017-5046-8
  [arXiv:1705.01848 [hep-th]].
  %%CITATION = doi:10.1140/epjc/s10052-017-5046-8;%%
  %22 citations counted in INSPIRE as of 15 Apr 2019

%\cite{Bosma:2019aiu}
\bibitem{Bosma:2019aiu} 
  L.~Bosma, B.~Knorr and F.~Saueressig,
  %``Resolving Spacetime Singularities within Quantum Gravity,''
  arXiv:1904.04845 [hep-th].
  %%CITATION = ARXIV:1904.04845;%%

%\cite{Afshordi:2015iza}
\bibitem{Afshordi:2015iza} 
  N.~Afshordi and E.~Nelson,
  %``Cosmological bounds on TeV-scale physics and beyond,''
  Phys.\ Rev.\ D {\bf 93}, no. 8, 083505 (2016)
  doi:10.1103/PhysRevD.93.083505
  [arXiv:1504.00012 [hep-th]].
  %%CITATION = doi:10.1103/PhysRevD.93.083505;%%
  %7 citations counted in INSPIRE as of 15 Apr 2019


%\cite{Bonanno:2001xi}
\bibitem{Bonanno:2001xi} 
  A.~Bonanno and M.~Reuter,
  %``Cosmology of the Planck era from a renormalization group for quantum gravity,''
  Phys.\ Rev.\ D {\bf 65}, 043508 (2002)
  doi:10.1103/PhysRevD.65.043508
  [hep-th/0106133].
  %%CITATION = doi:10.1103/PhysRevD.65.043508;%%
  %238 citations counted in INSPIRE as of 15 Apr 2019


%\cite{Reuter:2003ca}
\bibitem{Reuter:2003ca} 
  M.~Reuter and H.~Weyer,
  %``Renormalization group improved gravitational actions: A Brans-Dicke approach,''
  Phys.\ Rev.\ D {\bf 69}, 104022 (2004)
  doi:10.1103/PhysRevD.69.104022
  [hep-th/0311196].
  %%CITATION = doi:10.1103/PhysRevD.69.104022;%%
  %120 citations counted in INSPIRE as of 15 Apr 2019


%\cite{Reuter:2004nx}
\bibitem{Reuter:2004nx} 
  M.~Reuter and H.~Weyer,
  %``Quantum gravity at astrophysical distances?,''
  JCAP {\bf 0412}, 001 (2004)
  doi:10.1088/1475-7516/2004/12/001
  [hep-th/0410119].
  %%CITATION = doi:10.1088/1475-7516/2004/12/001;%%
  %154 citations counted in INSPIRE as of 15 Apr 2019


%\cite{Weinberg:2009wa}
\bibitem{Weinberg:2009wa} 
  S.~Weinberg,
  %``Asymptotically Safe Inflation,''
  Phys.\ Rev.\ D {\bf 81}, 083535 (2010)
  doi:10.1103/PhysRevD.81.083535
  [arXiv:0911.3165 [hep-th]].
  %%CITATION = doi:10.1103/PhysRevD.81.083535;%%
  %107 citations counted in INSPIRE as of 15 Apr 2019


%\cite{Tye:2010an}
\bibitem{Tye:2010an} 
  S.-H.~H.~Tye and J.~Xu,
  %``Comment on Asymptotically Safe Inflation,''
  Phys.\ Rev.\ D {\bf 82}, 127302 (2010)
  doi:10.1103/PhysRevD.82.127302
  [arXiv:1008.4787 [hep-th]].
  %%CITATION = doi:10.1103/PhysRevD.82.127302;%%
  %30 citations counted in INSPIRE as of 15 Apr 2019


%\cite{Bonanno:2010bt}
\bibitem{Bonanno:2010bt} 
  A.~Bonanno, A.~Contillo and R.~Percacci,
  %``Inflationary solutions in asymptotically safe f(R) theories,''
  Class.\ Quant.\ Grav.\  {\bf 28}, 145026 (2011)
  doi:10.1088/0264-9381/28/14/145026
  [arXiv:1006.0192 [gr-qc]].
  %%CITATION = doi:10.1088/0264-9381/28/14/145026;%%
  %77 citations counted in INSPIRE as of 15 Apr 2019


%\cite{Eichhorn:2010tb}
\bibitem{Eichhorn:2010tb} 
  A.~Eichhorn and H.~Gies,
  %``Ghost anomalous dimension in asymptotically safe quantum gravity,''
  Phys.\ Rev.\ D {\bf 81}, 104010 (2010)
  doi:10.1103/PhysRevD.81.104010
  [arXiv:1001.5033 [hep-th]].
  %%CITATION = doi:10.1103/PhysRevD.81.104010;%%
  %103 citations counted in INSPIRE as of 15 Apr 2019


%\cite{Litim:2011cp}
\bibitem{Litim:2011cp} 
  D.~F.~Litim,
  %``Renormalisation group and the Planck scale,''
  Phil.\ Trans.\ Roy.\ Soc.\ Lond.\ A {\bf 369}, 2759 (2011)
  doi:10.1098/rsta.2011.0103
  [arXiv:1102.4624 [hep-th]].
  %%CITATION = doi:10.1098/rsta.2011.0103;%%
  %145 citations counted in INSPIRE as of 15 Apr 2019


%\cite{Grande:2011xf}
\bibitem{Grande:2011xf} 
  J.~Grande, J.~Sola, S.~Basilakos and M.~Plionis,
  %``Hubble expansion and structure formation in the 'running FLRW model' of the cosmic evolution,''
  JCAP {\bf 1108}, 007 (2011)
  doi:10.1088/1475-7516/2011/08/007
  [arXiv:1103.4632 [astro-ph.CO]].
  %%CITATION = doi:10.1088/1475-7516/2011/08/007;%%
  %83 citations counted in INSPIRE as of 15 Apr 2019


%\cite{Benedetti:2012dx}
\bibitem{Benedetti:2012dx} 
  D.~Benedetti and F.~Caravelli,
  %``The Local potential approximation in quantum gravity,''
  JHEP {\bf 1206}, 017 (2012)
  Erratum: [JHEP {\bf 1210}, 157 (2012)]
  doi:10.1007/JHEP06(2012)017, 10.1007/JHEP10(2012)157
  [arXiv:1204.3541 [hep-th]].
  %%CITATION = doi:10.1007/JHEP06(2012)017, 10.1007/JHEP10(2012)157;%%
  %101 citations counted in INSPIRE as of 15 Apr 2019


%\cite{Hindmarsh:2012rc}
\bibitem{Hindmarsh:2012rc} 
  M.~Hindmarsh and I.~D.~Saltas,
  %``f(R) Gravity from the renormalisation group,''
  Phys.\ Rev.\ D {\bf 86}, 064029 (2012)
  doi:10.1103/PhysRevD.86.064029
  [arXiv:1203.3957 [gr-qc]].
  %%CITATION = doi:10.1103/PhysRevD.86.064029;%%
  %41 citations counted in INSPIRE as of 15 Apr 2019


%\cite{Eichhorn:2012va}
\bibitem{Eichhorn:2012va} 
  A.~Eichhorn,
  %``Quantum-gravity-induced matter self-interactions in the asymptotic-safety scenario,''
  Phys.\ Rev.\ D {\bf 86}, 105021 (2012)
  doi:10.1103/PhysRevD.86.105021
  [arXiv:1204.0965 [gr-qc]].
  %%CITATION = doi:10.1103/PhysRevD.86.105021;%%
  %60 citations counted in INSPIRE as of 15 Apr 2019


%\cite{Dona:2013qba}
\bibitem{Dona:2013qba} 
  P.~Don\'a, A.~Eichhorn and R.~Percacci,
  %``Matter matters in asymptotically safe quantum gravity,''
  Phys.\ Rev.\ D {\bf 89}, no. 8, 084035 (2014)
  doi:10.1103/PhysRevD.89.084035
  [arXiv:1311.2898 [hep-th]].
  %%CITATION = doi:10.1103/PhysRevD.89.084035;%%
  %124 citations counted in INSPIRE as of 15 Apr 2019


%\cite{Copeland:2013vva}
\bibitem{Copeland:2013vva} 
  E.~J.~Copeland, C.~Rahmede and I.~D.~Saltas,
  %``Asymptotically Safe Starobinsky Inflation,''
  Phys.\ Rev.\ D {\bf 91}, no. 10, 103530 (2015)
  doi:10.1103/PhysRevD.91.103530
  [arXiv:1311.0881 [gr-qc]].
  %%CITATION = doi:10.1103/PhysRevD.91.103530;%%
  %54 citations counted in INSPIRE as of 15 Apr 2019


%\cite{Bonanno:2015fga}
\bibitem{Bonanno:2015fga} 
  A.~Bonanno and A.~Platania,
  %``Asymptotically safe inflation from quadratic gravity,''
  Phys.\ Lett.\ B {\bf 750}, 638 (2015)
  doi:10.1016/j.physletb.2015.10.005
  [arXiv:1507.03375 [gr-qc]].
  %%CITATION = doi:10.1016/j.physletb.2015.10.005;%%
  %32 citations counted in INSPIRE as of 15 Apr 2019


%\cite{Rodrigues:2015hba}
\bibitem{Rodrigues:2015hba} 
  D.~C.~Rodrigues, B.~Chauvineau and O.~F.~Piattella,
  %``Scalar-Tensor gravity with system-dependent potential and its relation with Renormalization Group extended General Relativity,''
  JCAP {\bf 1509}, no. 09, 009 (2015)
  doi:10.1088/1475-7516/2015/09/009
  [arXiv:1504.05119 [gr-qc]].
  %%CITATION = doi:10.1088/1475-7516/2015/09/009;%%
  %19 citations counted in INSPIRE as of 15 Apr 2019


%\cite{Modesto:2015ozb}
\bibitem{Modesto:2015ozb} 
  L.~Modesto and I.~L.~Shapiro,
  %``Superrenormalizable quantum gravity with complex ghosts,''
  Phys.\ Lett.\ B {\bf 755}, 279 (2016)
  doi:10.1016/j.physletb.2016.02.021
  [arXiv:1512.07600 [hep-th]].
  %%CITATION = doi:10.1016/j.physletb.2016.02.021;%%
  %60 citations counted in INSPIRE as of 15 Apr 2019


%\cite{Rodrigues:2015rya}
\bibitem{Rodrigues:2015rya} 
  D.~C.~Rodrigues, B.~Koch, O.~F.~Piattella and I.~L.~Shapiro,
  %``The bending of light within gravity with large scale renormalization group effects,''
  AIP Conf.\ Proc.\  {\bf 1647}, 57 (2015).
  doi:10.1063/1.4913338
  %%CITATION = doi:10.1063/1.4913338;%%
  %6 citations counted in INSPIRE as of 15 Apr 2019


%\cite{Rodrigues:2016tfm}
\bibitem{Rodrigues:2016tfm} 
  D.~C.~Rodrigues, S.~Mauro and \'A.~O.~F.~de Almeida,
  %``Solar System constraints on Renormalization Group extended General Relativity: The PPN and Laplace-Runge-Lenz analyses with the external potential effect,''
  Phys.\ Rev.\ D {\bf 94}, no. 8, 084036 (2016)
  doi:10.1103/PhysRevD.94.084036
  [arXiv:1609.03613 [gr-qc]].
  %%CITATION = doi:10.1103/PhysRevD.94.084036;%%
  %6 citations counted in INSPIRE as of 15 Apr 2019


%\cite{Koch:2016uso}
\bibitem{Koch:2016uso} 
  B.~Koch, I.~A.~Reyes and \'A.~Rinc\'on,
  %``A scale-dependent black hole in three-dimensional space–time,''
  Class.\ Quant.\ Grav.\  {\bf 33}, no. 22, 225010 (2016)
  doi:10.1088/0264-9381/33/22/225010
  [arXiv:1606.04123 [hep-th]].
  %%CITATION = doi:10.1088/0264-9381/33/22/225010;%%
  %24 citations counted in INSPIRE as of 15 Apr 2019


%\cite{Platania:2017djo}
\bibitem{Platania:2017djo} 
  A.~Platania and F.~Saueressig,
  %``Functional Renormalization Group Flows on Friedman–Lemaître–Robertson–Walker backgrounds,''
  Found.\ Phys.\  {\bf 48}, no. 10, 1291 (2018)
  doi:10.1007/s10701-018-0181-0
  [arXiv:1710.01972 [hep-th]].
  %%CITATION = doi:10.1007/s10701-018-0181-0;%%
  %6 citations counted in INSPIRE as of 15 Apr 2019


%\cite{Bonanno:2017gji}
\bibitem{Bonanno:2017gji} 
  A.~Bonanno, S.~J.~Gabriele Gionti and A.~Platania,
  %``Bouncing and emergent cosmologies from Arnowitt–Deser–Misner RG flows,''
  Class.\ Quant.\ Grav.\  {\bf 35}, no. 6, 065004 (2018)
  doi:10.1088/1361-6382/aaa535
  [arXiv:1710.06317 [gr-qc]].
  %%CITATION = doi:10.1088/1361-6382/aaa535;%%
  %12 citations counted in INSPIRE as of 15 Apr 2019


%\cite{Rincon:2017ypd}
\bibitem{Rincon:2017ypd} 
  \'A.~Rinc\'on, B.~Koch and I.~Reyes,
  %``BTZ black hole assuming running couplings,''
  J.\ Phys.\ Conf.\ Ser.\  {\bf 831}, no. 1, 012007 (2017)
  doi:10.1088/1742-6596/831/1/012007
  [arXiv:1701.04531 [hep-th]].
  %%CITATION = doi:10.1088/1742-6596/831/1/012007;%%
  %24 citations counted in INSPIRE as of 15 Apr 2019


%\cite{Rincon:2017goj}
\bibitem{Rincon:2017goj} 
  \'A.~Rinc\'on, E.~Contreras, P.~Bargue\~no, B.~Koch, G.~Panotopoulos and A.~Hern\'andez-Arboleda,
  %``scale-dependent three-dimensional charged black holes in linear and non-linear electrodynamics,''
  Eur.\ Phys.\ J.\ C {\bf 77}, no. 7, 494 (2017)
  doi:10.1140/epjc/s10052-017-5045-9
  [arXiv:1704.04845 [hep-th]].
  %%CITATION = doi:10.1140/epjc/s10052-017-5045-9;%%
  %26 citations counted in INSPIRE as of 15 Apr 2019


%\cite{Modesto:2017hzl}
\bibitem{Modesto:2017hzl} 
  L.~Modesto, L.~Rachwal and I.~L.~Shapiro,
  %``Renormalization group in super-renormalizable quantum gravity,''
  Eur.\ Phys.\ J.\ C {\bf 78}, no. 7, 555 (2018)
  doi:10.1140/epjc/s10052-018-6035-2
  [arXiv:1704.03988 [hep-th]].
  %%CITATION = doi:10.1140/epjc/s10052-018-6035-2;%%
  %16 citations counted in INSPIRE as of 15 Apr 2019


%\cite{Contreras:2017eza}
\bibitem{Contreras:2017eza} 
  E.~Contreras, \'A.~Rinc\'on, B.~Koch and P.~Bargue\~no,
  %``A regular scale-dependent black hole solution,''
  Int.\ J.\ Mod.\ Phys.\ D {\bf 27}, no. 03, 1850032 (2017)
  doi:10.1142/S0218271818500323
  [arXiv:1711.08400 [gr-qc]].
  %%CITATION = doi:10.1142/S0218271818500323;%%
  %21 citations counted in INSPIRE as of 15 Apr 2019


%\cite{Calmet:2017rxl}
\bibitem{Calmet:2017rxl} 
  X.~Calmet, S.~Capozziello and D.~Pryer,
  %``Gravitational Effective Action at Second Order in Curvature and Gravitational Waves,''
  Eur.\ Phys.\ J.\ C {\bf 77}, no. 9, 589 (2017)
  doi:10.1140/epjc/s10052-017-5172-3
  [arXiv:1708.08253 [hep-th]].
  %%CITATION = doi:10.1140/epjc/s10052-017-5172-3;%%
  %13 citations counted in INSPIRE as of 15 Apr 2019


%\cite{Toniato:2017wmk}
\bibitem{Toniato:2017wmk} 
  J.~D.~Toniato, D.~C.~Rodrigues, \'A.~O.~F.~de Almeida and N.~Bertini,
  %``Will-Nordtvedt PPN formalism applied to renormalization group extensions of general relativity,''
  Phys.\ Rev.\ D {\bf 96}, no. 6, 064034 (2017)
  doi:10.1103/PhysRevD.96.064034
  [arXiv:1706.09032 [gr-qc]].
  %%CITATION = doi:10.1103/PhysRevD.96.064034;%%
  %2 citations counted in INSPIRE as of 15 Apr 2019


%\cite{Merzlikin:2017zan}
\bibitem{Merzlikin:2017zan} 
  B.~S.~Merzlikin, I.~L.~Shapiro, A.~Wipf and O.~Zanusso,
  %``Renormalization group flows and fixed points for a scalar field in curved space with nonminimal $F(\phi)R$ coupling,''
  Phys.\ Rev.\ D {\bf 96}, no. 12, 125007 (2017)
  doi:10.1103/PhysRevD.96.125007
  [arXiv:1711.02224 [hep-th]].
  %%CITATION = doi:10.1103/PhysRevD.96.125007;%%
  %3 citations counted in INSPIRE as of 15 Apr 2019


%\cite{Rincon:2018sgd}
\bibitem{Rincon:2018sgd} 
  \'A.~Rinc\'on and G.~Panotopoulos,
  %``Quasinormal modes of scale-dependent black holes in ( 1+2 )-dimensional Einstein-power-Maxwell theory,''
  Phys.\ Rev.\ D {\bf 97}, no. 2, 024027 (2018)
  doi:10.1103/PhysRevD.97.024027
  [arXiv:1801.03248 [hep-th]].
  %%CITATION = doi:10.1103/PhysRevD.97.024027;%%
  %15 citations counted in INSPIRE as of 15 Apr 2019


%\cite{Hernandez-Arboleda:2018qdo}
\bibitem{Hernandez-Arboleda:2018qdo} 
  A.~Hern\'andez-Arboleda, \'A.~Rinc\'on, B.~Koch, E.~Contreras and P.~Bargue\~no,
  %``Preliminary test of cosmological models in the scale-dependent scenario,''
  arXiv:1802.05288 [gr-qc].
  %%CITATION = ARXIV:1802.05288;%%
  %10 citations counted in INSPIRE as of 15 Apr 2019


%\cite{Contreras:2018dhs}
\bibitem{Contreras:2018dhs} 
  E.~Contreras, \'A.~Rinc\'on, B.~Koch and P.~Bargue\~no,
  %``Scale-dependent polytropic black hole,''
  Eur.\ Phys.\ J.\ C {\bf 78}, no. 3, 246 (2018)
  doi:10.1140/epjc/s10052-018-5709-0
  [arXiv:1803.03255 [gr-qc]].
  %%CITATION = doi:10.1140/epjc/s10052-018-5709-0;%%
  %14 citations counted in INSPIRE as of 15 Apr 2019


%\cite{Rincon:2018lyd}
\bibitem{Rincon:2018lyd} 
  \'A.~Rinc\'on and B.~Koch,
  %``Scale-dependent BTZ black hole,''
  Eur.\ Phys.\ J.\ C {\bf 78}, no. 12, 1022 (2018)
  doi:10.1140/epjc/s10052-018-6488-3
  [arXiv:1806.03024 [hep-th]].
  %%CITATION = doi:10.1140/epjc/s10052-018-6488-3;%%
  %12 citations counted in INSPIRE as of 15 Apr 2019


%\cite{Rincon:2018dsq}
\bibitem{Rincon:2018dsq} 
  \'A.~Rinc\'on, E.~Contreras, P.~Bargue\~no, B.~Koch and G.~Panotopoulos,
  %``Scale-dependent ( $2+1$ )-dimensional electrically charged black holes in Einstein-power-Maxwell theory,''
  Eur.\ Phys.\ J.\ C {\bf 78}, no. 8, 641 (2018)
  doi:10.1140/epjc/s10052-018-6106-4
  [arXiv:1807.08047 [hep-th]].
  %%CITATION = doi:10.1140/epjc/s10052-018-6106-4;%%
  %11 citations counted in INSPIRE as of 15 Apr 2019


%\cite{Calmet:2018elv}
\bibitem{Calmet:2018elv} 
  X.~Calmet,
  %``Vanishing of Quantum Gravitational Corrections to Vacuum Solutions of General Relativity at Second Order in Curvature,''
  Phys.\ Lett.\ B {\bf 787}, 36 (2018)
  doi:10.1016/j.physletb.2018.10.040
  [arXiv:1810.09719 [hep-th]].
  %%CITATION = doi:10.1016/j.physletb.2018.10.040;%%
  %1 citations counted in INSPIRE as of 15 Apr 2019


%\cite{Contreras:2018gct}
\bibitem{Contreras:2018gct} 
  E.~Contreras, \'A.~Rinc\'on and J.~M.~Ram\'irez-Velasquez,
  %``Relativistic dust accretion onto a scale--dependent polytropic black hole,''
  Eur.\ Phys.\ J.\ C {\bf 79}, no. 1, 53 (2019)
  doi:10.1140/epjc/s10052-019-6601-2
  [arXiv:1810.07356 [gr-qc]].
  %%CITATION = doi:10.1140/epjc/s10052-019-6601-2;%%
  %6 citations counted in INSPIRE as of 15 Apr 2019


%\cite{Bonanno:2018gck}
\bibitem{Bonanno:2018gck} 
  A.~Bonanno, A.~Platania and F.~Saueressig,
  %``Cosmological bounds on the field content of asymptotically safe gravity–matter models,''
  Phys.\ Lett.\ B {\bf 784}, 229 (2018)
  doi:10.1016/j.physletb.2018.06.047
  [arXiv:1803.02355 [gr-qc]].
  %%CITATION = doi:10.1016/j.physletb.2018.06.047;%%
  %10 citations counted in INSPIRE as of 15 Apr 2019


%\cite{Contreras:2018gpl}
\bibitem{Contreras:2018gpl} 
  E.~Contreras and P.~Bargue\~no,
  %``Scale--dependent Hayward black hole and the generalized uncertainty principle,''
  Mod.\ Phys.\ Lett.\ A {\bf 33}, no. 32, 1850184 (2018)
  doi:10.1142/S0217732318501845
  [arXiv:1809.00785 [gr-qc]].
  %%CITATION = doi:10.1142/S0217732318501845;%%
  %7 citations counted in INSPIRE as of 15 Apr 2019


%\cite{Contreras:2018swc}
\bibitem{Contreras:2018swc} 
  E.~Contreras and P.~Bargue\~no,
  %``A self-sustained traversable scale-dependent wormhole,''
  Int.\ J.\ Mod.\ Phys.\ D {\bf 27}, no. 09, 1850101 (2018)
  doi:10.1142/S0218271818501018
  [arXiv:1804.00988 [gr-qc]].
  %%CITATION = doi:10.1142/S0218271818501018;%%
  %10 citations counted in INSPIRE as of 15 Apr 2019


%\cite{Sahni:2002kh}
\bibitem{Sahni:2002kh} 
  V.~Sahni,
  %``The Cosmological constant problem and quintessence,''
  Class.\ Quant.\ Grav.\  {\bf 19}, 3435 (2002)
  doi:10.1088/0264-9381/19/13/304
  [astro-ph/0202076].
  %%CITATION = doi:10.1088/0264-9381/19/13/304;%%
  %279 citations counted in INSPIRE as of 15 Apr 2019


%\cite{Prokopec:2006yh}
\bibitem{Prokopec:2006yh} 
  T.~Prokopec,
  %``A Solution to the cosmological constant problem,''
  gr-qc/0603088.
  %%CITATION = GR-QC/0603088;%%
  %18 citations counted in INSPIRE as of 15 Apr 2019


%\cite{Brodsky:2009zd}
\bibitem{Brodsky:2009zd} 
  S.~J.~Brodsky and R.~Shrock,
  %``Condensates in Quantum Chromodynamics and the Cosmological Constant,''
  Proc.\ Nat.\ Acad.\ Sci.\  {\bf 108}, 45 (2011)
  doi:10.1073/pnas.1010113107
  [arXiv:0905.1151 [hep-th]].
  %%CITATION = doi:10.1073/pnas.1010113107;%%
  %121 citations counted in INSPIRE as of 15 Apr 2019


%\cite{Kaloper:2013zca}
\bibitem{Kaloper:2013zca} 
  N.~Kaloper and A.~Padilla,
  %``Sequestering the Standard Model Vacuum Energy,''
  Phys.\ Rev.\ Lett.\  {\bf 112}, no. 9, 091304 (2014)
  doi:10.1103/PhysRevLett.112.091304
  [arXiv:1309.6562 [hep-th]].
  %%CITATION = doi:10.1103/PhysRevLett.112.091304;%%
  %95 citations counted in INSPIRE as of 15 Apr 2019


%\cite{Stojkovic:2014lha}
\bibitem{Stojkovic:2014lha} 
  D.~Stojkovic,
  %``Vanishing dimensions : A review,''
  Mod.\ Phys.\ Lett.\ A {\bf 28}, 1330034 (2013)
  doi:10.1142/S0217732313300346
  [arXiv:1406.2696 [gr-qc]].
  %%CITATION = doi:10.1142/S0217732313300346;%%
  %20 citations counted in INSPIRE as of 15 Apr 2019


%\cite{Padilla:2015aaa}
\bibitem{Padilla:2015aaa} 
  A.~Padilla,
  %``Lectures on the Cosmological Constant Problem,''
  arXiv:1502.05296 [hep-th].
  %%CITATION = ARXIV:1502.05296;%%
  %101 citations counted in INSPIRE as of 15 Apr 2019


%\cite{Bass:2015yaa}
\bibitem{Bass:2015yaa} 
  S.~D.~Bass,
  %``Vacuum energy and the cosmological constant,''
  Mod.\ Phys.\ Lett.\ A {\bf 30}, no. 22, 1540033 (2015)
  doi:10.1142/S0217732315400337
  [arXiv:1503.05483 [hep-ph]].
  %%CITATION = doi:10.1142/S0217732315400337;%%
  %8 citations counted in INSPIRE as of 15 Apr 2019


%\cite{Novikov:2016hrc}
\bibitem{Novikov:2016hrc} 
  E.~A.~Novikov,
  %``Ultralight gravitons with tiny electric dipole moment are seeping from the vacuum,''
  Mod.\ Phys.\ Lett.\ A {\bf 31}, no. 15, 1650092 (2016).
  doi:10.1142/S0217732316500929
  %%CITATION = doi:10.1142/S0217732316500929;%%
  %11 citations counted in INSPIRE as of 15 Apr 2019


%\cite{Nojiri:2016mlb}
\bibitem{Nojiri:2016mlb} 
  S.~Nojiri,
  %``Some solutions for one of the cosmological constant problems,''
  Mod.\ Phys.\ Lett.\ A {\bf 31}, no. 37, 1650213 (2016)
  doi:10.1142/S0217732316502138
  [arXiv:1601.02203 [hep-th]].
  %%CITATION = doi:10.1142/S0217732316502138;%%
  %20 citations counted in INSPIRE as of 15 Apr 2019


%\cite{Biemans:2016rvp}
\bibitem{Biemans:2016rvp} 
  J.~Biemans, A.~Platania and F.~Saueressig,
  %``Quantum gravity on foliated spacetimes: Asymptotically safe and sound,''
  Phys.\ Rev.\ D {\bf 95}, no. 8, 086013 (2017)
  doi:10.1103/PhysRevD.95.086013
  [arXiv:1609.04813 [hep-th]].
  %%CITATION = doi:10.1103/PhysRevD.95.086013;%%
  %48 citations counted in INSPIRE as of 30 Jul 2019


%\cite{Hossenfelder:2018ikr}
\bibitem{Hossenfelder:2018ikr} 
  S.~Hossenfelder,
  %``Screams for Explanation: Finetuning and Naturalness in the Foundations of Physics,''
  arXiv:1801.02176 [physics.hist-ph].
  %%CITATION = ARXIV:1801.02176;%%
  %9 citations counted in INSPIRE as of 15 Apr 2019


%\cite{Weinberg:1976xy}
\bibitem{Weinberg:1976xy} 
  S.~Weinberg,
  %``Critical Phenomena for Field Theorists,''
  doi:10.1007/978-1-4684-0931-4 1
  %%CITATION = doi:10.1007/978-1-4684-0931-4 1;%%
  %19 citations counted in INSPIRE as of 15 Apr 2019


%\cite{Hawking:1979ig}
\bibitem{Hawking:1979ig} 
  S.~W.~Hawking and W.~Israel,
  %``General Relativity : An Einstein Centenary Survey,''
  %%CITATION = INSPIRE-149363;%%
  %23 citations counted in INSPIRE as of 15 Apr 2019


%\cite{Wetterich:1992yh}
\bibitem{Wetterich:1992yh} 
  C.~Wetterich,
  %``Exact evolution equation for the effective potential,''
  Phys.\ Lett.\ B {\bf 301}, 90 (1993)
  doi:10.1016/0370-2693(93)90726-X
  [arXiv:1710.05815 [hep-th]].
  %%CITATION = doi:10.1016/0370-2693(93)90726-X;%%
  %1309 citations counted in INSPIRE as of 15 Apr 2019


%\cite{Morris:1993qb}
\bibitem{Morris:1993qb} 
  T.~R.~Morris,
  %``The Exact renormalization group and approximate solutions,''
  Int.\ J.\ Mod.\ Phys.\ A {\bf 9}, 2411 (1994)
  doi:10.1142/S0217751X94000972
  [hep-ph/9308265].
  %%CITATION = doi:10.1142/S0217751X94000972;%%
  %512 citations counted in INSPIRE as of 15 Apr 2019


%\cite{Bonanno:2000ep}
\bibitem{Bonanno:2000ep} 
  A.~Bonanno and M.~Reuter,
  %``Renormalization group improved black hole space-times,''
  Phys.\ Rev.\ D {\bf 62}, 043008 (2000)
  doi:10.1103/PhysRevD.62.043008
  [hep-th/0002196].
  %%CITATION = doi:10.1103/PhysRevD.62.043008;%%
  %294 citations counted in INSPIRE as of 15 Apr 2019


%\cite{Litim:2002xm}
\bibitem{Litim:2002xm} 
  D.~F.~Litim and J.~M.~Pawlowski,
  %``Completeness and consistency of renormalisation group flows,''
  Phys.\ Rev.\ D {\bf 66}, 025030 (2002)
  doi:10.1103/PhysRevD.66.025030
  [hep-th/0202188].
  %%CITATION = doi:10.1103/PhysRevD.66.025030;%%
  %110 citations counted in INSPIRE as of 15 Apr 2019


%\cite{Reuter:2004nv}
\bibitem{Reuter:2004nv} 
  M.~Reuter and H.~Weyer,
  %``Running Newton constant, improved gravitational actions, and galaxy rotation curves,''
  Phys.\ Rev.\ D {\bf 70}, 124028 (2004)
  doi:10.1103/PhysRevD.70.124028
  [hep-th/0410117].
  %%CITATION = doi:10.1103/PhysRevD.70.124028;%%
  %113 citations counted in INSPIRE as of 15 Apr 2019


%\cite{Bonanno:2006eu}
\bibitem{Bonanno:2006eu} 
  A.~Bonanno and M.~Reuter,
  %``Spacetime structure of an evaporating black hole in quantum gravity,''
  Phys.\ Rev.\ D {\bf 73}, 083005 (2006)
  doi:10.1103/PhysRevD.73.083005
  [hep-th/0602159].
  %%CITATION = doi:10.1103/PhysRevD.73.083005;%%
  %152 citations counted in INSPIRE as of 15 Apr 2019


%\cite{Niedermaier:2006ns}
\bibitem{Niedermaier:2006ns} 
  M.~Niedermaier,
  %``The Asymptotic safety scenario in quantum gravity: An Introduction,''
  Class.\ Quant.\ Grav.\  {\bf 24}, R171 (2007)
  doi:10.1088/0264-9381/24/18/R01
  [gr-qc/0610018].
  %%CITATION = doi:10.1088/0264-9381/24/18/R01;%%
  %205 citations counted in INSPIRE as of 15 Apr 2019


%\cite{Percacci:2007sz}
\bibitem{Percacci:2007sz} 
  R.~Percacci,
  %``Asymptotic Safety,''
  In *Oriti, D. (ed.): Approaches to quantum gravity* 111-128
  [arXiv:0709.3851 [hep-th]].
  %%CITATION = ARXIV:0709.3851;%%
  %238 citations counted in INSPIRE as of 15 Apr 2019


%\cite{Bonanno:1998ye}
\bibitem{Bonanno:1998ye} 
  A.~Bonanno and M.~Reuter,
  %``Quantum gravity effects near the null black hole singularity,''
  Phys.\ Rev.\ D {\bf 60}, 084011 (1999)
  doi:10.1103/PhysRevD.60.084011
  [gr-qc/9811026].
  %%CITATION = doi:10.1103/PhysRevD.60.084011;%%
  %104 citations counted in INSPIRE as of 15 Apr 2019


%\cite{Emoto:2005te}
\bibitem{Emoto:2005te} 
  H.~Emoto,
  %``Asymptotic safety of quantum gravity and improved spacetime of black hole singularity by cutoff identification,''
  hep-th/0511075.
  %%CITATION = HEP-TH/0511075;%%
  %10 citations counted in INSPIRE as of 15 Apr 2019


%\cite{Reuter:2006rg}
\bibitem{Reuter:2006rg} 
  M.~Reuter and E.~Tuiran,
  %``Quantum Gravity Effects in Rotating Black Holes,''
  doi:10.1142/9789812834300 0473
  hep-th/0612037.
  %%CITATION = doi:10.1142/9789812834300 0473;%%
  %40 citations counted in INSPIRE as of 15 Apr 2019


%\cite{Ward:2006vw}
\bibitem{Ward:2006vw} 
  B.~F.~L.~Ward,
  %``Planck Scale Remnants in Resummed Quantum Gravity,''
  Acta Phys.\ Polon.\ B {\bf 37}, 1967 (2006)
  [hep-ph/0605054].
  %%CITATION = HEP-PH/0605054;%%
  %25 citations counted in INSPIRE as of 15 Apr 2019


%\cite{Koch:2007yt}
\bibitem{Koch:2007yt} 
  B.~Koch,
  %``Renormalization group and black hole production in large extra dimensions,''
  Phys.\ Lett.\ B {\bf 663}, 334 (2008)
  doi:10.1016/j.physletb.2008.04.025
  [arXiv:0707.4644 [hep-ph]].
  %%CITATION = doi:10.1016/j.physletb.2008.04.025;%%
  %31 citations counted in INSPIRE as of 15 Apr 2019


%\cite{Hewett:2007st}
\bibitem{Hewett:2007st} 
  J.~Hewett and T.~Rizzo,
  %``Collider Signals of Gravitational Fixed Points,''
  JHEP {\bf 0712}, 009 (2007)
  doi:10.1088/1126-6708/2007/12/009
  [arXiv:0707.3182 [hep-ph]].
  %%CITATION = doi:10.1088/1126-6708/2007/12/009;%%
  %61 citations counted in INSPIRE as of 15 Apr 2019


%\cite{Litim:2007iu}
\bibitem{Litim:2007iu} 
  D.~F.~Litim and T.~Plehn,
  %``Signatures of gravitational fixed points at the LHC,''
  Phys.\ Rev.\ Lett.\  {\bf 100}, 131301 (2008)
  doi:10.1103/PhysRevLett.100.131301
  [arXiv:0707.3983 [hep-ph]].
  %%CITATION = doi:10.1103/PhysRevLett.100.131301;%%
  %69 citations counted in INSPIRE as of 15 Apr 2019


%\cite{Burschil:2009va}
\bibitem{Burschil:2009va} 
  T.~Burschil and B.~Koch,
  %``Renormalization group improved black hole space-time in large extra dimensions,''
  Zh.\ Eksp.\ Teor.\ Fiz.\  {\bf 92}, 219 (2010)
  [JETP Lett.\  {\bf 92}, 193 (2010)]
  doi:10.1134/S0021364010160010
  [arXiv:0912.4517 [hep-ph]].
  %%CITATION = doi:10.1134/S0021364010160010;%%
  %28 citations counted in INSPIRE as of 15 Apr 2019


%\cite{Falls:2010he}
\bibitem{Falls:2010he} 
  K.~Falls, D.~F.~Litim and A.~Raghuraman,
  %``Black Holes and Asymptotically Safe Gravity,''
  Int.\ J.\ Mod.\ Phys.\ A {\bf 27}, 1250019 (2012)
  doi:10.1142/S0217751X12500194
  [arXiv:1002.0260 [hep-th]].
  %%CITATION = doi:10.1142/S0217751X12500194;%%
  %75 citations counted in INSPIRE as of 15 Apr 2019


%\cite{Casadio:2010fw}
\bibitem{Casadio:2010fw} 
  R.~Casadio, S.~D.~H.~Hsu and B.~Mirza,
  %``Asymptotic Safety, Singularities, and Gravitational Collapse,''
  Phys.\ Lett.\ B {\bf 695}, 317 (2011)
  doi:10.1016/j.physletb.2010.10.060
  [arXiv:1008.2768 [gr-qc]].
  %%CITATION = doi:10.1016/j.physletb.2010.10.060;%%
  %28 citations counted in INSPIRE as of 15 Apr 2019


%\cite{Basu:2010nf}
\bibitem{Basu:2010nf} 
  S.~Basu and D.~Mattingly,
  %``Asymptotic Safety, Asymptotic Darkness, and the hoop conjecture in the extreme UV,''
  Phys.\ Rev.\ D {\bf 82}, 124017 (2010)
  doi:10.1103/PhysRevD.82.124017
  [arXiv:1006.0718 [hep-th]].
  %%CITATION = doi:10.1103/PhysRevD.82.124017;%%
  %21 citations counted in INSPIRE as of 15 Apr 2019


%\cite{Reuter:2010xb}
\bibitem{Reuter:2010xb} 
  M.~Reuter and E.~Tuiran,
  %``Quantum Gravity Effects in the Kerr Spacetime,''
  Phys.\ Rev.\ D {\bf 83}, 044041 (2011)
  doi:10.1103/PhysRevD.83.044041
  [arXiv:1009.3528 [hep-th]].
  %%CITATION = doi:10.1103/PhysRevD.83.044041;%%
  %43 citations counted in INSPIRE as of 15 Apr 2019


%\cite{Cai:2010zh}
\bibitem{Cai:2010zh} 
  Y.~F.~Cai and D.~A.~Easson,
  %``Black holes in an asymptotically safe gravity theory with higher derivatives,''
  JCAP {\bf 1009}, 002 (2010)
  doi:10.1088/1475-7516/2010/09/002
  [arXiv:1007.1317 [hep-th]].
  %%CITATION = doi:10.1088/1475-7516/2010/09/002;%%
  %47 citations counted in INSPIRE as of 15 Apr 2019


%\cite{Falls:2012nd}
\bibitem{Falls:2012nd} 
  K.~Falls and D.~F.~Litim,
  %``Black hole thermodynamics under the microscope,''
  Phys.\ Rev.\ D {\bf 89}, 084002 (2014)
  doi:10.1103/PhysRevD.89.084002
  [arXiv:1212.1821 [gr-qc]].
  %%CITATION = doi:10.1103/PhysRevD.89.084002;%%
  %54 citations counted in INSPIRE as of 15 Apr 2019


%\cite{Becker:2012js}
\bibitem{Becker:2012js} 
  D.~Becker and M.~Reuter,
  %``Running boundary actions, Asymptotic Safety, and black hole thermodynamics,''
  JHEP {\bf 1207}, 172 (2012)
  doi:10.1007/JHEP07(2012)172
  [arXiv:1205.3583 [hep-th]].
  %%CITATION = doi:10.1007/JHEP07(2012)172;%%
  %46 citations counted in INSPIRE as of 15 Apr 2019


%\cite{Becker:2012jx}
\bibitem{Becker:2012jx} 
  D.~Becker and M.~Reuter,
  %``Asymptotic Safety and Black Hole Thermodynamics,''
  doi:10.1142/9789814623995 0405
  arXiv:1212.4274 [hep-th].
  %%CITATION = doi:10.1142/9789814623995 0405;%%
  %20 citations counted in INSPIRE as of 15 Apr 2019


%\cite{Koch:2013owa}
\bibitem{Koch:2013owa} 
  B.~Koch and F.~Saueressig,
  %``Structural aspects of asymptotically safe black holes,''
  Class.\ Quant.\ Grav.\  {\bf 31}, 015006 (2014)
  doi:10.1088/0264-9381/31/1/015006
  [arXiv:1306.1546 [hep-th]].
  %%CITATION = doi:10.1088/0264-9381/31/1/015006;%%
  %46 citations counted in INSPIRE as of 15 Apr 2019


%\cite{Gonzalez:2015upa}
\bibitem{Gonzalez:2015upa} 
  C.~Gonz\'alez and B.~Koch,
  %``Improved Reissner–Nordström–(A)dS black hole in asymptotic safety,''
  Int.\ J.\ Mod.\ Phys.\ A {\bf 31}, no. 26, 1650141 (2016)
  doi:10.1142/S0217751X16501414
  [arXiv:1508.01502 [hep-th]].
  %%CITATION = doi:10.1142/S0217751X16501414;%%
  %6 citations counted in INSPIRE as of 15 Apr 2019


%\cite{Torres:2017ygl}
\bibitem{Torres:2017ygl} 
  R.~Torres,
  %``Nonsingular black holes, the cosmological constant, and asymptotic safety,''
  Phys.\ Rev.\ D {\bf 95}, no. 12, 124004 (2017)
  doi:10.1103/PhysRevD.95.124004
  [arXiv:1703.09997 [gr-qc]].
  %%CITATION = doi:10.1103/PhysRevD.95.124004;%%
  %8 citations counted in INSPIRE as of 15 Apr 2019


%\cite{Knorr:2017fus}
\bibitem{Knorr:2017fus} 
  B.~Knorr and S.~Lippoldt,
  %``Correlation functions on a curved background,''
  Phys.\ Rev.\ D {\bf 96}, no. 6, 065020 (2017)
  doi:10.1103/PhysRevD.96.065020
  [arXiv:1707.01397 [hep-th]].
  %%CITATION = doi:10.1103/PhysRevD.96.065020;%%
  %26 citations counted in INSPIRE as of 15 Apr 2019


%\cite{Eichhorn:2017egq}
\bibitem{Eichhorn:2017egq} 
  A.~Eichhorn,
  %``Status of the asymptotic safety paradigm for quantum gravity and matter,''
  Found.\ Phys.\  {\bf 48}, no. 10, 1407 (2018)
  doi:10.1007/s10701-018-0196-6
  [arXiv:1709.03696 [gr-qc]].
  %%CITATION = doi:10.1007/s10701-018-0196-6;%%
  %42 citations counted in INSPIRE as of 15 Apr 2019


%\cite{Pawlowski:2018swz}
\bibitem{Pawlowski:2018swz} 
  J.~M.~Pawlowski and D.~Stock,
  %``Quantum-improved Schwarzschild-(A)dS and Kerr-(A)dS spacetimes,''
  Phys.\ Rev.\ D {\bf 98}, no. 10, 106008 (2018)
  doi:10.1103/PhysRevD.98.106008
  [arXiv:1807.10512 [hep-th]].
  %%CITATION = doi:10.1103/PhysRevD.98.106008;%%
  %7 citations counted in INSPIRE as of 15 Apr 2019


%\cite{Alkofer:2018zze}
\bibitem{Alkofer:2018zze} 
  N.~Alkofer,
  %``Quantum Gravity from Fundamental Questions to Phenomenological Applications,''
  arXiv:1810.03132 [hep-th].
  %%CITATION = ARXIV:1810.03132;%%
  %2 citations counted in INSPIRE as of 15 Apr 2019


%\cite{Eichhorn:2018yfc}
\bibitem{Eichhorn:2018yfc} 
  A.~Eichhorn,
  %``An asymptotically safe guide to quantum gravity and matter,''
  arXiv:1810.07615 [hep-th].
  %%CITATION = ARXIV:1810.07615;%%
  %21 citations counted in INSPIRE as of 15 Apr 2019


%\cite{Koch:2010nn}
\bibitem{Koch:2010nn} 
  B.~Koch and I.~Ramirez,
  %``Exact renormalization group with optimal scale and its application to cosmology,''
  Class.\ Quant.\ Grav.\  {\bf 28}, 055008 (2011)
  doi:10.1088/0264-9381/28/5/055008
  [arXiv:1010.2799 [gr-qc]].
  %%CITATION = doi:10.1088/0264-9381/28/5/055008;%%
  %46 citations counted in INSPIRE as of 15 Apr 2019


%\cite{Domazet:2012tw}
\bibitem{Domazet:2012tw} 
  S.~Domazet and H.~Stefancic,
  %``Renormalization group scale-setting from the action - a road to modified gravity theories,''
  Class.\ Quant.\ Grav.\  {\bf 29}, 235005 (2012)
  doi:10.1088/0264-9381/29/23/235005
  [arXiv:1204.1483 [gr-qc]].
  %%CITATION = doi:10.1088/0264-9381/29/23/235005;%%
  %16 citations counted in INSPIRE as of 15 Apr 2019


%\cite{Koch:2014joa}
\bibitem{Koch:2014joa} 
  B.~Koch, P.~Rioseco and C.~Contreras,
  %``Scale Setting for Self-consistent Backgrounds,''
  Phys.\ Rev.\ D {\bf 91}, no. 2, 025009 (2015)
  doi:10.1103/PhysRevD.91.025009
  [arXiv:1409.4443 [hep-th]].
  %%CITATION = doi:10.1103/PhysRevD.91.025009;%%
  %25 citations counted in INSPIRE as of 15 Apr 2019


%\cite{Manrique:2009uh}
\bibitem{Manrique:2009uh} 
  E.~Manrique and M.~Reuter,
  %``Bimetric Truncations for Quantum Einstein Gravity and Asymptotic Safety,''
  Annals Phys.\  {\bf 325}, 785 (2010)
  doi:10.1016/j.aop.2009.11.009
  [arXiv:0907.2617 [gr-qc]].
  %%CITATION = doi:10.1016/j.aop.2009.11.009;%%
  %122 citations counted in INSPIRE as of 15 Apr 2019


%\cite{Manrique:2010am}
\bibitem{Manrique:2010am} 
  E.~Manrique, M.~Reuter and F.~Saueressig,
  %``Bimetric Renormalization Group Flows in Quantum Einstein Gravity,''
  Annals Phys.\  {\bf 326}, 463 (2011)
  doi:10.1016/j.aop.2010.11.006
  [arXiv:1006.0099 [hep-th]].
  %%CITATION = doi:10.1016/j.aop.2010.11.006;%%
  %126 citations counted in INSPIRE as of 15 Apr 2019


%\cite{Manrique:2010mq}
\bibitem{Manrique:2010mq} 
  E.~Manrique, M.~Reuter and F.~Saueressig,
  %``Matter Induced Bimetric Actions for Gravity,''
  Annals Phys.\  {\bf 326}, 440 (2011)
  doi:10.1016/j.aop.2010.11.003
  [arXiv:1003.5129 [hep-th]].
  %%CITATION = doi:10.1016/j.aop.2010.11.003;%%
  %98 citations counted in INSPIRE as of 15 Apr 2019


%\cite{Becker:2014qya}
\bibitem{Becker:2014qya} 
  D.~Becker and M.~Reuter,
  %``En route to Background Independence: Broken split-symmetry, and how to restore it with bi-metric average actions,''
  Annals Phys.\  {\bf 350}, 225 (2014)
  doi:10.1016/j.aop.2014.07.023
  [arXiv:1404.4537 [hep-th]].
  %%CITATION = doi:10.1016/j.aop.2014.07.023;%%
  %96 citations counted in INSPIRE as of 15 Apr 2019


%\cite{Dietz:2015owa}
\bibitem{Dietz:2015owa} 
  J.~A.~Dietz and T.~R.~Morris,
  %``Background independent exact renormalization group for conformally reduced gravity,''
  JHEP {\bf 1504}, 118 (2015)
  doi:10.1007/JHEP04(2015)118
  [arXiv:1502.07396 [hep-th]].
  %%CITATION = doi:10.1007/JHEP04(2015)118;%%
  %51 citations counted in INSPIRE as of 15 Apr 2019


%\cite{Labus:2016lkh}
\bibitem{Labus:2016lkh} 
  P.~Labus, T.~R.~Morris and Z.~H.~Slade,
  %``Background independence in a background dependent renormalization group,''
  Phys.\ Rev.\ D {\bf 94}, no. 2, 024007 (2016)
  doi:10.1103/PhysRevD.94.024007
  [arXiv:1603.04772 [hep-th]].
  %%CITATION = doi:10.1103/PhysRevD.94.024007;%%
  %44 citations counted in INSPIRE as of 15 Apr 2019


%\cite{Morris:2016spn}
\bibitem{Morris:2016spn} 
  T.~R.~Morris,
  %``Large curvature and background scale independence in single-metric approximations to asymptotic safety,''
  JHEP {\bf 1611}, 160 (2016)
  doi:10.1007/JHEP11(2016)160
  [arXiv:1610.03081 [hep-th]].
  %%CITATION = doi:10.1007/JHEP11(2016)160;%%
  %31 citations counted in INSPIRE as of 15 Apr 2019


%\cite{Ohta:2017dsq}
\bibitem{Ohta:2017dsq} 
  N.~Ohta,
  %``Background Scale Independence in Quantum Gravity,''
  PTEP {\bf 2017}, no. 3, 033E02 (2017)
  doi:10.1093/ptep/ptx020
  [arXiv:1701.01506 [hep-th]].
  %%CITATION = doi:10.1093/ptep/ptx020;%%
  %18 citations counted in INSPIRE as of 15 Apr 2019

%\cite{Koch:2014cqa}
\bibitem{Koch:2014cqa} 
  B.~Koch and F.~Saueressig,
  %``Black holes within Asymptotic Safety,''
  Int.\ J.\ Mod.\ Phys.\ A {\bf 29}, no. 8, 1430011 (2014)
  doi:10.1142/S0217751X14300117
  [arXiv:1401.4452 [hep-th]].
  %%CITATION = doi:10.1142/S0217751X14300117;%%
  %43 citations counted in INSPIRE as of 29 Jul 2019

%\cite{Rincon:2017ayr}
\bibitem{Rincon:2017ayr} 
  \'A.~Rinc\'on and B.~Koch,
  %``On the null energy condition in scale-dependent frameworks with spherical symmetry,''
  J.\ Phys.\ Conf.\ Ser.\  {\bf 1043}, no. 1, 012015 (2018)
  doi:10.1088/1742-6596/1043/1/012015
  [arXiv:1705.02729 [hep-th]].
  %%CITATION = doi:10.1088/1742-6596/1043/1/012015;%%
  %18 citations counted in INSPIRE as of 15 Apr 2019


%\cite{Jacobson:2007tj}
\bibitem{Jacobson:2007tj} 
  T.~Jacobson,
  %``When is g(tt) g(rr) = -1?,''
  Class.\ Quant.\ Grav.\  {\bf 24}, 5717 (2007)
  doi:10.1088/0264-9381/24/22/N02
  [arXiv:0707.3222 [gr-qc]].
  %%CITATION = doi:10.1088/0264-9381/24/22/N02;%%
  %57 citations counted in INSPIRE as of 15 Apr 2019


%\cite{Visser:1999de}
\bibitem{Visser:1999de} 
  M.~Visser and C.~Barcelo,
  %``Energy conditions and their cosmological implications,''
  doi:10.1142/9789812792129 0014
  gr-qc/0001099.
  %%CITATION = doi:10.1142/9789812792129 0014;%%
  %80 citations counted in INSPIRE as of 15 Apr 2019


%\cite{Canales:2018tbn}
\bibitem{Canales:2018tbn} 
  F.~Canales, B.~Koch, C.~Laporte and \'A.~Rinc\'on,
  %``Vacuum energy density: deflation during inflation,''
  arXiv:1812.10526 [gr-qc].
  %%CITATION = ARXIV:1812.10526;%%
  %4 citations counted in INSPIRE as of 15 Apr 2019


%\cite{Patrignani:2016xqp}
\bibitem{Patrignani:2016xqp} 
  C.~Patrignani {\it et al.} [Particle Data Group],
  %``Review of Particle Physics,''
  Chin.\ Phys.\ C {\bf 40}, no. 10, 100001 (2016).
  doi:10.1088/1674-1137/40/10/100001
  %%CITATION = doi:10.1088/1674-1137/40/10/100001;%%
  %4790 citations counted in INSPIRE as of 15 Apr 2019


%\cite{Banks:2003pt}
\bibitem{Banks:2003pt} 
  T.~Banks and W.~Fischler,
  %``An Upper bound on the number of e-foldings,''
  astro-ph/0307459.
  %%CITATION = ASTRO-PH/0307459;%%
  %65 citations counted in INSPIRE as of 15 Apr 2019


%\cite{Kolb:1990vq}
\bibitem{Kolb:1990vq} 
  E.~W.~Kolb and M.~S.~Turner,
  %``The Early Universe,''
  Front.\ Phys.\  {\bf 69}, 1 (1990).
  %%CITATION = FRPHA,69,1;%%
  %1576 citations counted in INSPIRE as of 15 Apr 2019


%\cite{Capozziello:1996xq}
\bibitem{Capozziello:1996xq} 
  S.~Capozziello, R.~de Ritis and A.~A.~Marino,
  %``A time-dependent 'cosmological constant' phenomenology,''
  Nuovo Cim.\ B {\bf 112}, 1351 (1997)
  [astro-ph/9605176].
  %%CITATION = ASTRO-PH/9605176;%%
  %9 citations counted in INSPIRE as of 15 Apr 2019


%\cite{Calmet:2001nu}
\bibitem{Calmet:2001nu} 
  X.~Calmet and H.~Fritzsch,
  %``The Cosmological evolution of the nucleon mass and the electroweak coupling constants,''
  Eur.\ Phys.\ J.\ C {\bf 24}, 639 (2002)
  doi:10.1007/s10052-002-0976-0
  [hep-ph/0112110].
  %%CITATION = doi:10.1007/s10052-002-0976-0;%%
  %174 citations counted in INSPIRE as of 15 Apr 2019


%\cite{Carneiro:2003as}
\bibitem{Carneiro:2003as} 
  S.~Carneiro,
  %``Decaying $\Lambda$ cosmology with varying G,''
  doi:10.1142/9789812704030 0204
  gr-qc/0307114.
  %%CITATION = doi:10.1142/9789812704030 0204;%%
  %12 citations counted in INSPIRE as of 15 Apr 2019


%\cite{Reisenegger:2009cq}
\bibitem{Reisenegger:2009cq} 
  A.~Reisenegger, P.~Jofr\'e and R.~Fern\'andez,
  %``Constraining a possible time-variation of the gravitational constant through “gravitochemical heating” of neutron stars,''
  Mem.\ Soc.\ Ast.\ It.\  {\bf 80}, no. 4, 829 (2009)
  [Highlights Astron.\  {\bf 15}, 314 (2010)]
  doi:10.1017/S1743921310009518
  [arXiv:0911.0190 [astro-ph.HE]].
  %%CITATION = doi:10.1017/S1743921310009518;%%
  %4 citations counted in INSPIRE as of 15 Apr 2019


%\cite{Yunes:2009bv}
\bibitem{Yunes:2009bv} 
  N.~Yunes, F.~Pretorius and D.~Spergel,
  %``Constraining the evolutionary history of Newton's constant with gravitational wave observations,''
  Phys.\ Rev.\ D {\bf 81}, 064018 (2010)
  doi:10.1103/PhysRevD.81.064018
  [arXiv:0912.2724 [gr-qc]].
  %%CITATION = doi:10.1103/PhysRevD.81.064018;%%
  %42 citations counted in INSPIRE as of 15 Apr 2019


%\cite{Uzan:2010pm}
\bibitem{Uzan:2010pm} 
  J.~P.~Uzan,
  %``Varying Constants, Gravitation and Cosmology,''
  Living Rev.\ Rel.\  {\bf 14}, 2 (2011)
  doi:10.12942/lrr-2011-2
  [arXiv:1009.5514 [astro-ph.CO]].
  %%CITATION = doi:10.12942/lrr-2011-2;%%
  %288 citations counted in INSPIRE as of 15 Apr 2019


%\cite{Fritzsch:2010zzc}
\bibitem{Fritzsch:2010zzc} 
  H.~Fritzsch,
  %``The fundamental constants in physics and their possible time variation,''
  Nucl.\ Phys.\ Proc.\ Suppl.\  {\bf 203-204}, 3 (2010).
  doi:10.1016/j.nuclphysbps.2010.08.002
  %%CITATION = doi:10.1016/j.nuclphysbps.2010.08.002;%%
  %2 citations counted in INSPIRE as of 15 Apr 2019


%\cite{Anderson:2015bva}
\bibitem{Anderson:2015bva} 
  J.~D.~Anderson, G.~Schubert, V.~Trimble and M.~R.~Feldman,
  %``Measurements of Newton's gravitational constant and the length of day,''
  EPL {\bf 110}, no. 1, 10002 (2015)
  doi:10.1209/0295-5075/110/10002
  [arXiv:1504.06604 [gr-qc]].
  %%CITATION = doi:10.1209/0295-5075/110/10002;%%
  %18 citations counted in INSPIRE as of 15 Apr 2019


%\cite{Kantha:2016ylw}
\bibitem{Kantha:2016ylw} 
  L.~Kantha,
  %``A Time-Dependent and Cosmological Model Consistent with Cosmological Constraints,''
  Adv.\ Astron.\  {\bf 2016}, 9743970 (2016).
  doi:10.1155/2016/9743970
  %%CITATION = doi:10.1155/2016/9743970;%%
  %1 citations counted in INSPIRE as of 15 Apr 2019


%\cite{Calmet:2017czo}
\bibitem{Calmet:2017czo} 
  X.~Calmet,
  %``Cosmological evolution of the Higgs boson’s vacuum expectation value,''
  Eur.\ Phys.\ J.\ C {\bf 77}, no. 11, 729 (2017)
  doi:10.1140/epjc/s10052-017-5324-5
  [arXiv:1707.06922 [gr-qc]].
  %%CITATION = doi:10.1140/epjc/s10052-017-5324-5;%%
  %2 citations counted in INSPIRE as of 15 Apr 2019


%\cite{Martin:1997ns}
\bibitem{Martin:1997ns} 
  S.~P.~Martin,
  %``A Supersymmetry primer,''
  Adv.\ Ser.\ Direct.\ High Energy Phys.\  {\bf 21}, 1 (2010)
  [Adv.\ Ser.\ Direct.\ High Energy Phys.\  {\bf 18}, 1 (1998)]
  doi:10.1142/9789812839657 0001, 10.1142/9789814307505 0001
  [hep-ph/9709356].
  %%CITATION = doi:10.1142/9789812839657 0001, 10.1142/9789814307505 0001;%%
  %3454 citations counted in INSPIRE as of 15 Apr 2019


%\cite{ArkaniHamed:1998rs}
\bibitem{ArkaniHamed:1998rs} 
  N.~Arkani-Hamed, S.~Dimopoulos and G.~R.~Dvali,
  %``The Hierarchy problem and new dimensions at a millimeter,''
  Phys.\ Lett.\ B {\bf 429}, 263 (1998)
  doi:10.1016/S0370-2693(98)00466-3
  [hep-ph/9803315].
  %%CITATION = doi:10.1016/S0370-2693(98)00466-3;%%
  %6449 citations counted in INSPIRE as of 15 Apr 2019


%\cite{Meissner:2006zh}
\bibitem{Meissner:2006zh} 
  K.~A.~Meissner and H.~Nicolai,
  %``Conformal Symmetry and the Standard Model,''
  Phys.\ Lett.\ B {\bf 648}, 312 (2007)
  doi:10.1016/j.physletb.2007.03.023
  [hep-th/0612165].
  %%CITATION = doi:10.1016/j.physletb.2007.03.023;%%
  %288 citations counted in INSPIRE as of 15 Apr 2019


%\cite{Reuter:1996cp}
\bibitem{Reuter:1996cp} 
  M.~Reuter,
  %``Nonperturbative evolution equation for quantum gravity,''
  Phys.\ Rev.\ D {\bf 57}, 971 (1998)
  doi:10.1103/PhysRevD.57.971
  [hep-th/9605030].
  %%CITATION = doi:10.1103/PhysRevD.57.971;%%
  %678 citations counted in INSPIRE as of 15 Apr 2019


%\cite{Reuter:2001ag}
\bibitem{Reuter:2001ag} 
  M.~Reuter and F.~Saueressig,
  %``Renormalization group flow of quantum gravity in the Einstein-Hilbert truncation,''
  Phys.\ Rev.\ D {\bf 65}, 065016 (2002)
  doi:10.1103/PhysRevD.65.065016
  [hep-th/0110054].
  %%CITATION = doi:10.1103/PhysRevD.65.065016;%%
  %345 citations counted in INSPIRE as of 15 Apr 2019


%\cite{Litim:2003vp}
\bibitem{Litim:2003vp} 
  D.~F.~Litim,
  %``Fixed points of quantum gravity,''
  Phys.\ Rev.\ Lett.\  {\bf 92}, 201301 (2004)
  doi:10.1103/PhysRevLett.92.201301
  [hep-th/0312114].
  %%CITATION = doi:10.1103/PhysRevLett.92.201301;%%
  %380 citations counted in INSPIRE as of 15 Apr 2019

\end{thebibliography}
\end{document}